\documentclass[secnumarabic, graphics,floatfix, nofootinbib,tightenlines,nobibnotes, aps, prl, 12pt]{revtex4-1}
\usepackage{graphicx}
\usepackage[english]{babel}
\usepackage[utf8]{inputenc}
\usepackage{amsmath,amssymb}
\usepackage{amsfonts}
\usepackage{multirow}
\usepackage{pstricks}
\usepackage{float}
\usepackage{subfigure}
\usepackage{color}
\usepackage{epsfig}
\usepackage{pst-tools}
\usepackage{mnsymbol}
\usepackage[section]{placeins}
\usepackage{geometry}
\def\be{\begin{equation}}
\def\ee{\end{equation}}
\def\bea{\begin{eqnarray}}
\def\eea{\end{eqnarray}}

\newcommand{\tabincell}[2]{\begin{tabular}{@{}#1@{}}#2\end{tabular}}

\begin{document}

\title{On nonlinearity in hydrodynamic response to the initial geometry in relativistic heavy-ion collisions}

\author{Dan Wen$^{1,2}$}
\author{Kai Lin$^{3,4}$}
\author{Wei-Liang Qian$^{4,2,1}$}
\author{Bin Wang$^{1}$}
\author{Yogiro Hama$^{5}$}
\author{Takeshi Kodama$^{6,7}$}

\affiliation{$^{1}$ School of Physical Science and Technology, Yangzhou University, 225002, Yangzhou, Jiangsu, China}
\affiliation{$^{2}$ Faculdade de Engenharia de Guaratinguet\'a, Universidade Estadual Paulista, 12516-410, Guaratinguet\'a, SP, Brazil}
\affiliation{$^{3}$ Hubei Subsurface Multi-scale Imaging Key Laboratory, Institute of Geophysics and Geomatics, China University of Geosciences, 430074, Wuhan, Hubei, China}
\affiliation{$^{4}$ Escola de Engenharia de Lorena, Universidade de S\~ao Paulo, 12602-810, Lorena, SP, Brazil}
\affiliation{$^{5}$ Instituto de F\'isica, Universidade de S\~ao Paulo, C.P. 66318, 05315-970, S\~ao Paulo, SP, Brazil}
\affiliation{$^{6}$ Instituto de F\'isica, Universidade Federal do Rio de Janeiro, C.P. 68528, 21945-970, Rio de Janeiro, RJ , Brazil}
\affiliation{$^{7}$ Instituto de F\'isica, Universidade Federal Fluminense, 24210-346, Niter\'oi, RJ, Brazil}

\date{Mar. 23, 2020}

\begin{abstract}

In the context of event-by-event hydrodynamic description, we analyze the implications of two models characterized by distinct initial conditions.
The initial energy density of the first model adopts a Gaussian-type distribution, while those of the second one are features by high energy peripheral tubes.
We calibrate the initial conditions of both models so that their initial probability distribution of eccentricity are mostly identical.
Subsequently, the resultant scaled probability distributions of collective flow and the correlations between flow harmonic and eccentricity coefficients are investigated.
Besides, the calculations are carried out for particle correlations regarding the symmetric cumulant, mixed harmonics, and nonlinear response coefficients.
Although the resultant two-particle correlations possess similar shapes, numerical calculations indicate a subtle difference between the two models.
To be specific, the difference resides in more detailed observables such as the probability distributions of elliptic flow as well as Pearson correlation coefficient regarding higher-order harmonics.
We discuss several essential aspects concerning the linearity and nonlinearity between initial eccentricities and final state anisotropies.
Further implications are addressed.


\end{abstract}

\maketitle

\section{I. Introduction}

The success of the hydrodynamic description of relativistic heavy-ion collisions plays a vital part in our ongoing endeavor to understand the properties of QCD matter~\cite{hydro-review-04,hydro-review-05,hydro-review-06,hydro-review-08,hydro-review-09,hydro-review-10}.
The essence of hydrodynamical evolution, by and large, has been attributed to the dynamic response to fluctuating initial conditions (IC). 
Moreover, as hydrodynamics is known for its highly nonlinear characteristics, various studies have been carried out to explore this aspect.
In particular, much efforts have been devoted to the relationship between initial state eccentricities and final state anisotropies~\cite{hydro-v3-02,hydro-vn-03,sph-vn-03,hydro-vn-04,sph-vn-04,sph-vn-06,hydro-vn-10}.

A quantitative notion on the decomposition of the anisotropic IC was first proposed in Refs.~\cite{hydro-v3-02,hydro-vn-03}.
The key idea of the study is that anisotropy of the IC can be decomposed in terms of a cumulant expansion, where the resulting expansion coefficients correspond to the ``connected" part of the eccentricity at a given order.
Therefore, a higher-order cumulant, by definition, has the contributions from the ``disconnected" combinations of the lower orders ones subtracted.
Moreover, flow harmonics are understood as the hydrodynamic response to IC fluctuations classified in terms of cumulants, while the lowest cumulants are assumed to have dominant effects.
In literature, for a given flow harmonic order, the contribution proportional to the cumulant of the same azimuthal order is attributed as much to the linear response.
While those proportional to the combinations of lower-order cumulants, which give rise to the same azimuthal order, are referred to as the nonlinear response.
In practice, it is noted that the response strength from different cumulant combinations is different.
Therefore, in practice, the ``best estimator" is taken to minimize the deviation from the perfect correlation~\cite{sph-vn-03,sph-vn-06}.
To be more specific, the established mapping between IC and flow harmonics resides in the correlation between an optimized linear combination of a given set of cumulant products and the corresponding flow harmonics.
By numerical studies have shown that such a mapping is indeed effective, particularly for the most central collisions.
These works have incited further efforts concerning this train of thought~\cite{hydro-vn-04,hydro-vn-10}.
As an example, it might be interested in verifying whether each individual component of a given azimuthal order indeed leads to a linear response by explicitly separating them from the IC~\cite{sph-vn-04}.
We note that, in this context, a given azimuthal harmonic order actually corresponds to an infinite set of moments or cumulants.
It is not entirely clear whether a specific one-to-one mapping shall exist, or in other words, to what extent different terms associated with the same azimuthal order will mix under dynamical evolution.
As a result, it still leaves room for other possibilities regarding the radial expansion.
In literatue, it has also been speculated that cumulant expansion might not provide the most ideal parametrization of initial conditions.
For instance, the Bessel-Fourier expansion proposed in Ref.~\cite{hydro-vn-ph-09,hydro-vn-ph-10} provides an exciting alternative in terms of orthonormal basis.
The advantage of the proposed IC decomposition resides in that fluctuations are ordered with respect to their wavelength.
Moreover, a radial mode related to a shorter wavelength can be suppressed more easily, and different modes might be mainly treated independently in a hydrodynamical formalism~\cite{hydro-vn-ph-11}. 

Another direction of approach is associated with flow analysis, and particularly concerning the higher harmonics and particle correlations.
Symmetric cumulant was proposed in Ref.~\cite{hydro-corr-ph-06} as a distinct observable tailored for the correlations between flow harmonics.
In particular, the symmetric cumulant does not depend on any particular event plane, neither on the correlation between them.
Moreover, it vanishes if the fluctuations of different flow harmonics are independent.
In this context, it is an excellent observable which is exclusively dedicated to exploring the correlations between the flow harmonics and their fluctuations.
As a comparison, event plane correlations can be studied by using the method proposed in Ref.~\cite{hydro-corr-ph-07}.
In fact, most of the above observables can be formally expressed in terms of the moments of flow, as discussed in Ref.~\cite{hydro-corr-ph-08}.
Here, the definitions of other quantities are derived through that of the complex anisotropic flow coefficient of $n$th harmonics $V_n$, namely, the Fourier coefficient of one particle distribution.
By evaluating the Pearson correlation coefficients between moments and other appropriately chosen quantities, one obtains the desired flow fluctuations, symmetric cumulant, and event plane correlations.  
More recently, the nonlinear response regarding ratios of mixed higher-order harmonic moments has been investigated by several authors.
Numerical studies are carried out in terms of transport as well as hydrodynamic models while the results are compared against the data~\cite{hydro-corr-ph-08,hydro-corr-04,hydro-corr-05}.
The ratio of the event average of the products of anisotropic flow coefficient, subsequently, give rise to various observables such as $v_n\{\Psi_m\}$, event planes correlation measured by CMS and ATLAS Collaborations~\cite{LHC-cms-vn-3,LHC-atlas-vn-3}.
Furthermore, according to the spirit of IC Fourier decomposition, the flow harmonics are divided into linear and nonlinear parts.
The analyzes are entirely based on flow harmonics, not directly related to IC eccentricities.
The linear response is attributed to IC fluctuations while the nonlinear part is to the mean geometric eccentricity. 
To separate the linear and nonlinear decompositions, in particular, the linear responses are assumed to be uncorrelated to the nonlinear ones.
The latter can be, again, expressed in terms of the ratios of the event average of the products of anisotropic flow coefficients.
Hydrodynamic simulations show that the corresponding results are comparable to the data.
Last but not least, principal component analysis (PCA)~\cite{zml-pca-01} has recently been employed by many authors for the flow analysis~\cite{hydro-vn-pca-01,hydro-vn-pca-02,hydro-vn-pca-03}.
PCA is a method widely applied in data analysis, inclusively for machine learning, which attempts to reorganize a complex data set into components expressed in most relevant dimensions.
As a result, the data can be presented in a lower-dimensional space.
In this context, when PCA applies, it significantly simplifies the underlying data structure.
Recently, the method is introduced to carry out analysis of the data on heavy-ion collisions~\cite{hydro-vn-pca-01}.
The main procedure of the PCA is to diagonalize a covariance matrix, where the covariance is evaluated with respect to two distinct types measurements, and the dimension of the matrix presents the total number of different features being measured.
In practice, the coordinates are translated conveniently in such a way that expected value is zero.
For the application of heavy-ion collisions, the covariance is calculated for the anisotropic flow coefficients evaluated at two different values of a chosen kinematic variable.
Thus the dimension of the matrix is taken to be the number of bins of the given quantity, for instance, transverse momentum or rapidity.
If the detector is azimuthally symmetric, the event average of $V_n$ vanishes, and therefore the numerical procedure can be readily implemented.
The validity of the application of PCA in the context of heavy-ion collisions resides in the fact that the dominant component turns out to be much more significant than others.
The main advantage of the method, as claimed by the authors, is that it makes use of all the information contained in the particle distribution function.
Nonetheless, the robustness of the method is being investigated~\cite{hydro-vn-pca-04}.

The above prominent methods for decomposition of the IC and flow harmonics have been extensively employed to explore relevant information regarding the collectivity of the system.
However, due to the nonlinearity of hydrodynamics, to what degree the mapping between IC and flow harmonics can be established quantitatively and therefore captured by the proposed methods is still not entirely settled.
Also, from the AdS/CFT viewpoint, hydrodynamics stands on the other side where the system is strongly interacting with intricated correlations, while according to the duality, the linear response theory is valid only for its dual gravity theory.
In this context, it is meaningful also to investigate some alternative approaches regarding the description of IC and its subsequent effect on collective phenomena, such as the Bessel-Fourier expansion mentioned above.
Recently, we proposed a peripheral tube model~\cite{sph-corr-02,sph-corr-03}, which provides an intuitive explanation for the generation of the triangular flow and two-particle correlations.
The model can also be viewed as an alternative approach in the context of event-by-event hydrodynamics. 
The IC is divided into background and fluctuations.
The former is obtained by averaging the distribution over many events from the same centrality class.
While for the latter, the IC fluctuations are understood as consisting of independent high energy spot located close to the surface of the system, referred to as peripheral tubes due to its longitudinal extension.
The resultant higher-order harmonics are attributed to how a peripheral tube affects the hydrodynamical evolution of the system locally.
The overall contributions are obtained by the superposition of those of individual tubes. 
The essential feature of our model is that the above picture attempts to interpret the IC fluctuations in terms of the localized ones, instead of the global sinusoidal expansion regarding moments.
To be specific, if a tube locates deep inside the hot matter, which might, however, contribute significantly to the moment expansion, is less relevant in our approach.
To the best of our knowledge, the effect of its hydrodynamic expansion would be mostly suppressed by its surroundings, causing less significant consequence in the media.
This is contrary to a tube staying close to the surface, which might cause significant disturbance to the one-particle azimuthal distribution, as well as the related two-particle correlations.
The model has been employed to study various features of the observed two-particle correlations in comparison with data~\cite{sph-corr-02,sph-corr-03,sph-corr-04,sph-corr-07,sph-corr-08}.

In this context, the present study is motivated to carry out a more close comparison between the IC of the peripheral tube model and those related to moment decomposition.
The primary strategy is to prepare two sets of event-by-event fluctuating IC with mostly identical eccentricity distributions and then investigate the subsequent linear as well as nonlinear hydrodynamic response and the resultant flow harmonics.
Although similar in terms of its Fourier components, the IC in question are visually distinct by construction.  
By employed most of the methods of IC and flow analysis mentioned above, we evaluated various relevant observables.
The differences between the two models are presented and discussed.
Furthermore, the implications of the present findings are addressed.

The present work is organized as follows.
In the next section, we briefly review the peripheral tube model and device an anisotropic Gaussian model which primarily consists of moments of Fourier decompositions.
The latter is mostly Guassin in the radial direction, and in the azimuthal direction, it is parameterized to contain different harmonic orders.
Subsequently, in Section III, we explain how the parameter of the anisotropic Gaussian IC are adjusted so that it gives largely identical eccentricity distributions to those of the peripheral tube model.
Both models are then fed to the hydrodynamic code SPheRIO.
The calculations are carried out for flow harmonics and its probability distributions, symmetric cumulant, mixed harmonics, as well as nonlinear response coefficients.
The results are presented in Section IV.
The last section is devoted to discussions and concluding remarks.

\section{II. The models}

In this section, we discuss the two models employed in the present study.
First, the main characteristics of the peripheral tube model are briefly summarized.
Then we devise an anisotropic Gaussian model, whose IC is tailored to reproduce the probability distribution of the eccentricities of the former.  

As mentioned above, the IC of the peripheral tube model consist of a smoothed background and a few high energy tubes close to the surface of the system.
The background gives rise to the averaged bulk properties of the system, while the tubes characterize the event-by-event fluctuations.
Subsequently, the energy density profile of the model is given by
\begin{eqnarray}
\epsilon = \epsilon_\mathrm{bgd}+\epsilon_\mathrm{tube} .
\end{eqnarray}
Here, the averaged background distribution reads
\begin{eqnarray}
\epsilon_\mathrm{bgd} = (K+Lr^2+Mr^4)e^{-r^{2c}}  \, , \label{avaIC} 
\end{eqnarray}
with
\begin{eqnarray}
r = \sqrt{ax^2+by^2} , 
\end{eqnarray}
where the parameters $K,~L,~M,~a,~b,~c$ are determined by a numerical fit to the averaged IC of Pb+Pb collisions for the $20\%-25\%$ centrality class at $2.76$ TeV, generated by EPOS~\cite{nexus-1,nexus-rept,epos-1,epos-2,epos-3}.
The profile of a high energy tube is given by 
\begin{eqnarray}
\epsilon_\mathrm{tube}&=& A_\mathrm{tube}\exp\left[{-\frac{(x-x_\mathrm{tube})^2-(y-y_\mathrm{tube})^2}{R_\mathrm{tube}}}\right] \, , \label{energytube}
\end{eqnarray}
with 
\begin{eqnarray}
r_\mathrm{tube}&=& \frac{r_0}{\sqrt{a\cos^2\theta+b\sin^2\theta}}   \\
x_\mathrm{tube}&=& r_\mathrm{tube}\cos(\theta)   \nonumber\\
y_\mathrm{tube}&=& r_\mathrm{tube}\sin(\theta)  \nonumber
\end{eqnarray}
where $A_\mathrm{tube}$ and $R_\mathrm{tube}$ are the maximum energy and radius of the tube, while $r_\mathrm{tube}$ give the radial location of the tube, subsequently determined by $r_0, a, b$, and $\theta$. 
The azimuth angle $\theta$ is randomly chosen for an individual tube.
In the present study, we will focus ourselves on IC with three randomly generated peripheral tubes.
The parameters used in the present study are summarized in Tab.~\ref{tubeparameters}.

\begin{table}[h!]
\begin{center}
  \caption{The model parameters of the peripheral tube model employed in the present study.}
  \begin{tabular}{|p{2cm}|p{2cm}|p{2cm}|}  
  \hline
    $K$ & $L$ & $M$ \\ 
   \hline
   103.9 & -89 & 28.5 \\
   \hline
   $a$ & $b$ & $c$ \\
   \hline
   0.077 & 0.033 & 2 \\
   \hline
   $A_\mathrm{tube}$ & $r_0$ & $R_\mathrm{tube}$ \\
   \hline
   30 & 1.3 & 1.1 \\  
  \hline  
  \end{tabular}
  \label{tubeparameters}
\end{center}
\end{table}

For the purpose of the present study, we introduce the following parameterization for an anisotropic Gaussian model.
\begin{eqnarray}
\epsilon(r,\theta) &=& Ze^{-\frac{r^2}{R^2(\theta)}} , \\
R(\theta) &=& R_0\left[1+\sum_{n=2}C_n\cos(n(\theta-\theta_n))\right]^{1/2} . \label{gaussianIC}
\end{eqnarray} 
Here, for a given azimuthal direction, the radial distribution is essentially Gaussian.
The azimuthal dependence of the radius is contained within the specific form of $R(\theta)$.
The latter draws a closed curve as one varies the azimuthal angle $\theta$ from $0$ to $2\pi$.
The value of the parameters $R_0$ and $Z$ are chosen accordingly so that its size and total energy 
\begin{eqnarray}
E_T=\int_0^{2\pi} d\theta \int_0^\infty dr\sqrt{-g} r \epsilon(r,\theta) = \pi \tau_0 R_0^2 Z
\end{eqnarray} 
are reminiscent to those of the averaged EPOS IC.
Here, $C_n$ and $\theta_n$ are randomized accordingly to reproduce the same eccentricity distribution of the tube model, as will be further discussed below.
The parameters employed for the anisotropic Gaussian model are summarized in the Tab.~\ref{Gaussianparameters}.
To be specific, the parameters $C_n$ are randomly chosen to satisfy a normal distribution centered at $M_n$ with standard deviation $\sigma_n$. 
This is carried out numerically using the Box-Muller method as follows.
One first picks out $U_1$ and $U_2$, two independent, uniformly distributed random numbers in the interval $[0,1]$.
Then, $C_n$ is evaluated accordingly to the following expressions
\begin{eqnarray}
G&=&\sqrt{-2\ln U_1}\cos(2\pi U_2) , \\
C_n&=& G\cdot \sigma_n + M_n .
\end{eqnarray}
If one obtains a negative $C_n$, it is simply be cast out.

\begin{table}[h!]
\begin{center}
  \caption{The model parameters of the anisotropic Gaussian model employed in the present study.}
  \begin{tabular}{|p{2cm}|p{2cm}|p{2cm}|}
  \hline 
    $R_0$ & \multicolumn{2}{|c|}{3.1} \\
   \hline
    $Z$   & \multicolumn{2}{|c|}{133} \\
   \hline
     $n$    & $\sigma_n$    &    $M_n$            \\
   \hline
     $2$      & 0.075 &  0.39         \\
   \hline
     $3$      & 0.095 &  0.045        \\
   \hline
     $4$      & 0.145 &  0.073        \\
   \hline
     $5$      & 0.128 &  0.063         \\     
  \hline  
  \end{tabular}
  \label{Gaussianparameters}
\end{center}
\end{table}

The eccentricities for a given IC of the anisotropic Gaussian model can be readily derived by evaluating the moments $\langle r^m\rangle$, $\langle r^m\cos m\theta\rangle$, and $\langle r^m\sin m\theta\rangle$, which turn out to be
\begin{eqnarray}
\langle r^m\rangle &=& \frac{R_0^m}{2\pi}\Gamma \left(\frac{m}{2}+1\right) I_m ,\\
\langle r^m\cos m\theta\rangle &=& \frac{R_0^m}{2\pi}\Gamma \left(\frac{m}{2}+1\right) I_m^C ,\nonumber\\
\langle r^m\sin m\theta\rangle &=& \frac{R_0^m}{2\pi}\Gamma \left(\frac{m}{2}+1\right) I_m^S ,\nonumber
\end{eqnarray} 
and therefore,
\begin{eqnarray}
\varepsilon_n &=& \frac{\sqrt{\langle r^n\cos(n\theta)\rangle^2 +\langle r^n\sin(n\theta)\rangle^2}}{\langle r^n\rangle}  = \frac{\sqrt{(I_m^C)^2+(I_m^S)^2}}{I_m} , \label{espilon}
\end{eqnarray} 
where
\begin{eqnarray}
I_m &=& \int_{0}^{2\pi} \left(1+\sum_{n=2}C_n\cos n(\theta-\theta_n)\right)^{\frac{m}{2}+1}d\theta ,\\
I_m^C &=& \int_{0}^{2\pi} \cos(m\theta)\left(1+\sum_{n=2}C_n\cos n(\theta-\theta_n)\right)^{\frac{m}{2}+1}d\theta ,\nonumber\\
I_m^S &=& \int_{0}^{2\pi} \sin(m\theta)\left(1+\sum_{n=2}C_n\cos n(\theta-\theta_n)\right)^{\frac{m}{2}+1}d\theta .\nonumber
\end{eqnarray} 
For small angular inhomogeneities, namely, $ Cn \ll 1$, one finds the following simplified expressions for the eccentricities
\begin{eqnarray}
\varepsilon_n \sim \frac{n+2}{4}C_n ,
\end{eqnarray}
and eccentricity planes
\begin{eqnarray}
\Phi_n \sim \frac{1}{n}\mathrm{arctan2}\frac{I_n^S}{I_n^C}+\frac{\pi}{n} .
\end{eqnarray}
In practice, numerical integrations are employed for the calculations.

\section{III. Numerical results}

\begin{figure}
\begin{tabular}{cc}
\vspace{-50pt}
\begin{minipage}{225pt}
\centerline{\includegraphics[width=350pt]{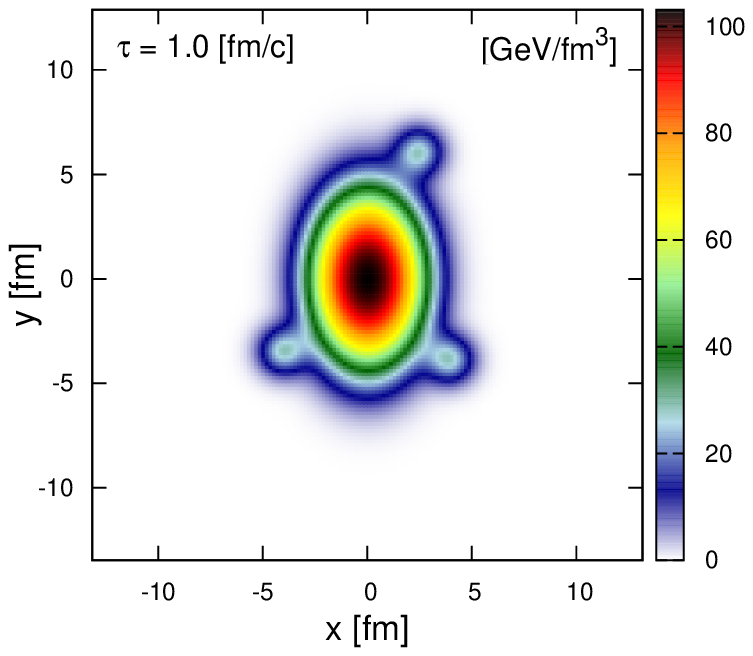}}
\end{minipage}
&
\begin{minipage}{225pt}
\centerline{\includegraphics[width=350pt]{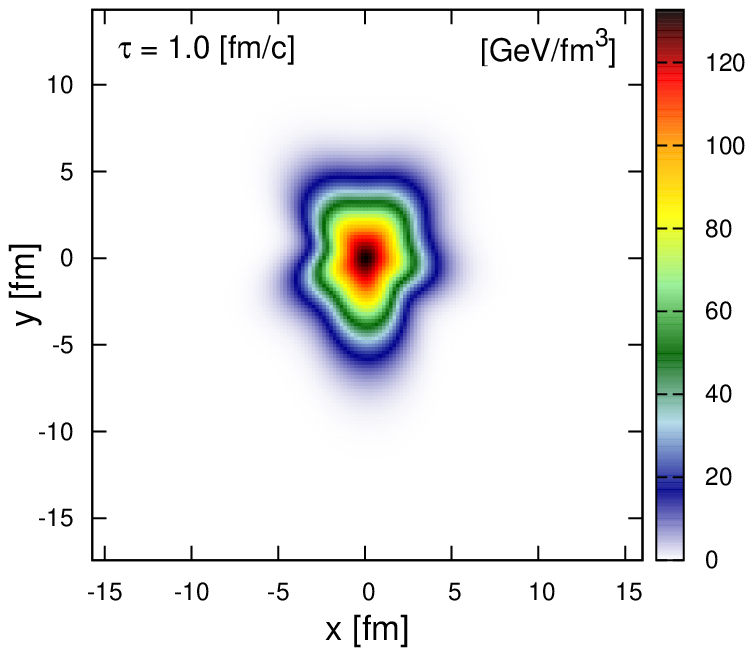}}
\end{minipage}
\\
\vspace{-50pt}
\begin{minipage}{225pt}
\centerline{\includegraphics[width=350pt]{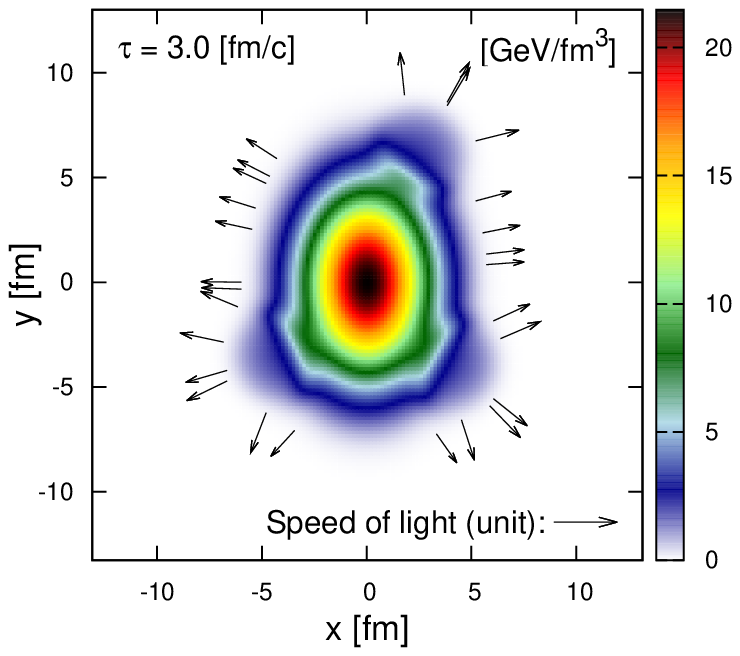}}
\end{minipage}
&
\begin{minipage}{225pt}
\centerline{\includegraphics[width=350pt]{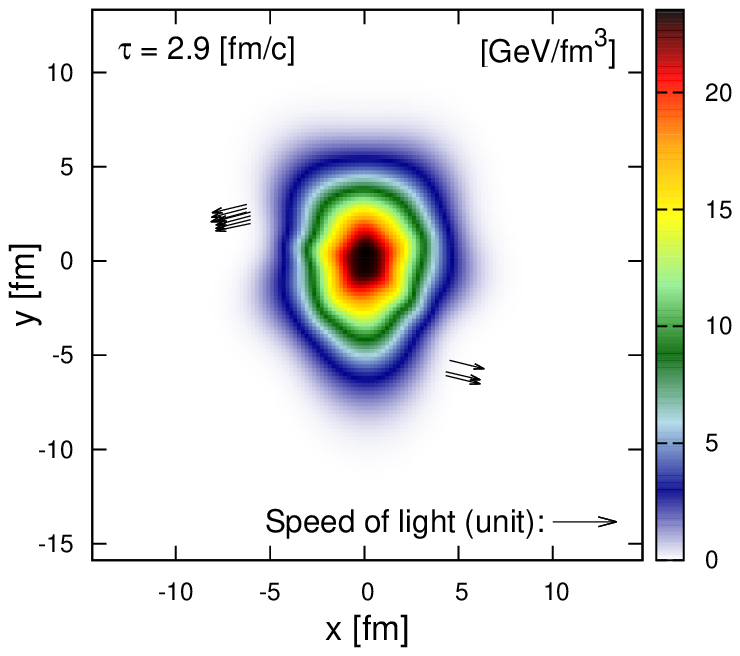}}
\end{minipage}
\\
\begin{minipage}{225pt}
\centerline{\includegraphics[width=350pt]{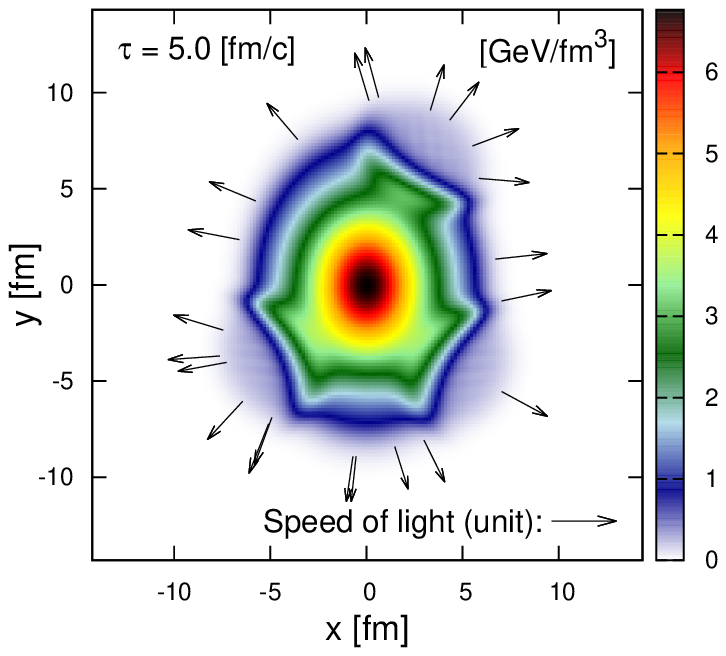}}
\end{minipage}
&
\begin{minipage}{225pt}
\centerline{\includegraphics[width=350pt]{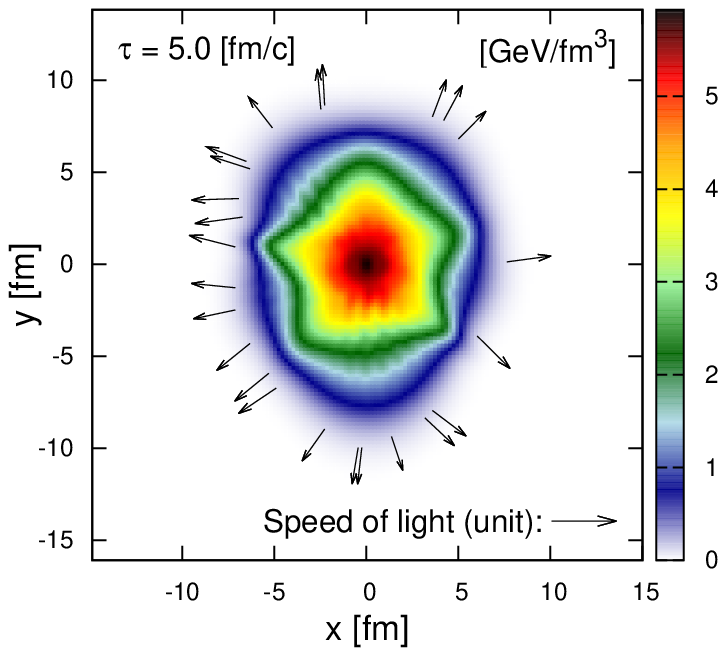}}
\end{minipage}
\end{tabular}
\renewcommand{\figurename}{Fig.}
\caption{(Color online)
The calculated temporal evolution of two random events of the peripheral tube model with three tubes (left column) and of anisotropic Gaussian model (right column).}
\label{evolutiontubej5}
\end{figure}

As discussed above, the parameterization of the peripheral tube model is adjusted to mimic event-by-event fluctuating initial conditions generated by EPOS.
On the other hand, those of the anisotropic Gaussian model is tailored accordingly to produce mostly identical eccentricity distribution.
However, both the IC and the subsequent temporal evolutions are quite different visually, as clearly demonstrated in Fig.~\ref{evolutiontubej5}. 
In the case of the tube model, evolution is evidently dominated by the deflection of the flow by the high energy tubes.
To be specific, the resultant peaks of particle emission are clearly associated with the locations of the three tubes, as discussed in Refs.~\cite{sph-corr-03,sph-corr-08}.
On the other hand, in the case of anisotropic Gaussian model, the overall energy distribution is more smooth.
It is rather difficult to predict the resultant evolution.
Although, on an event-by-event average, the apparent mapping between IC eccentricities and flow harmonics can be established, as discussed below.

\begin{figure}[htbp]
\begin{center}
\begin{minipage}{225pt}
\centering\includegraphics[width=220pt]{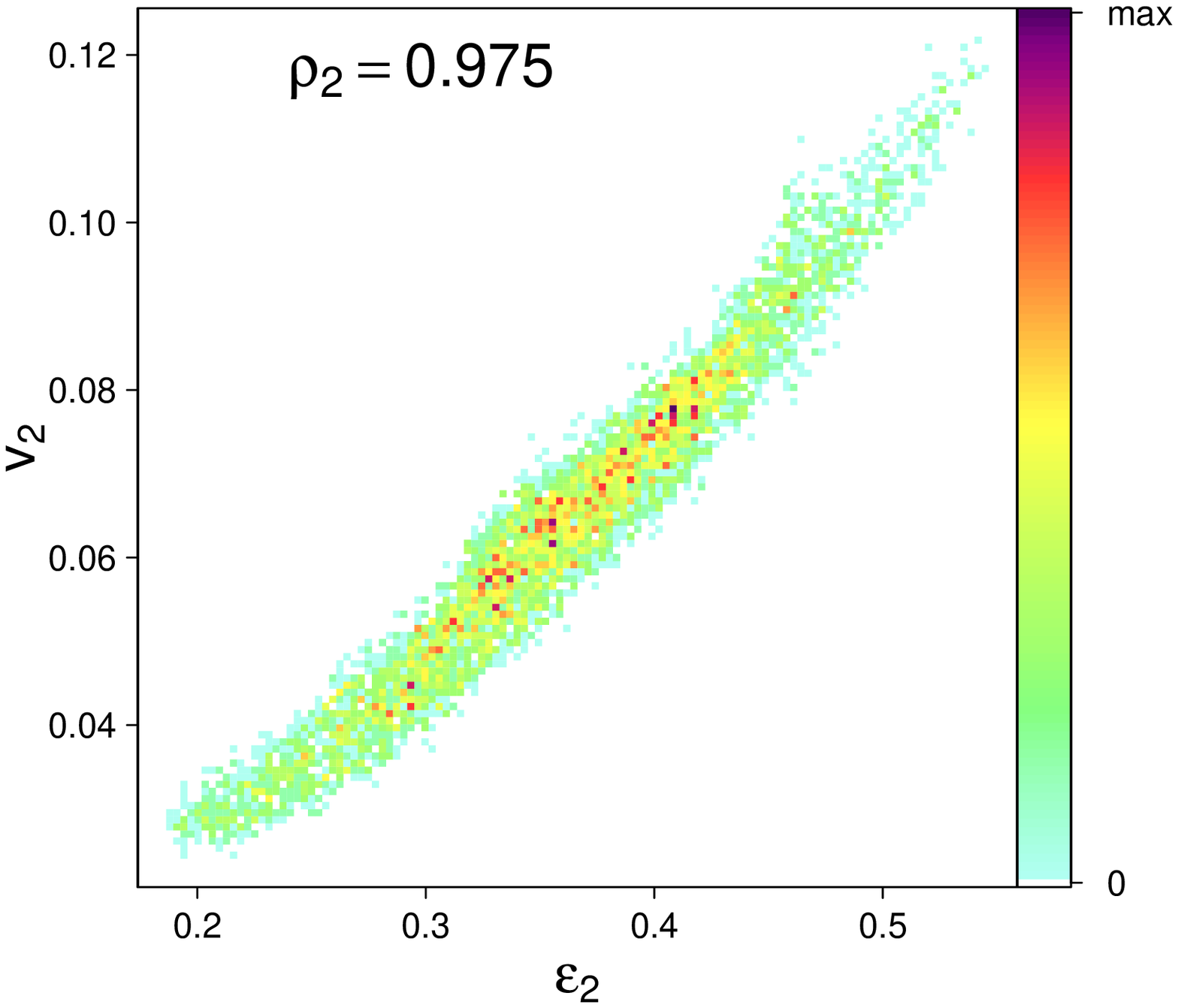}
\end{minipage}%
\begin{minipage}{225pt}
\centering\includegraphics[width=220pt]{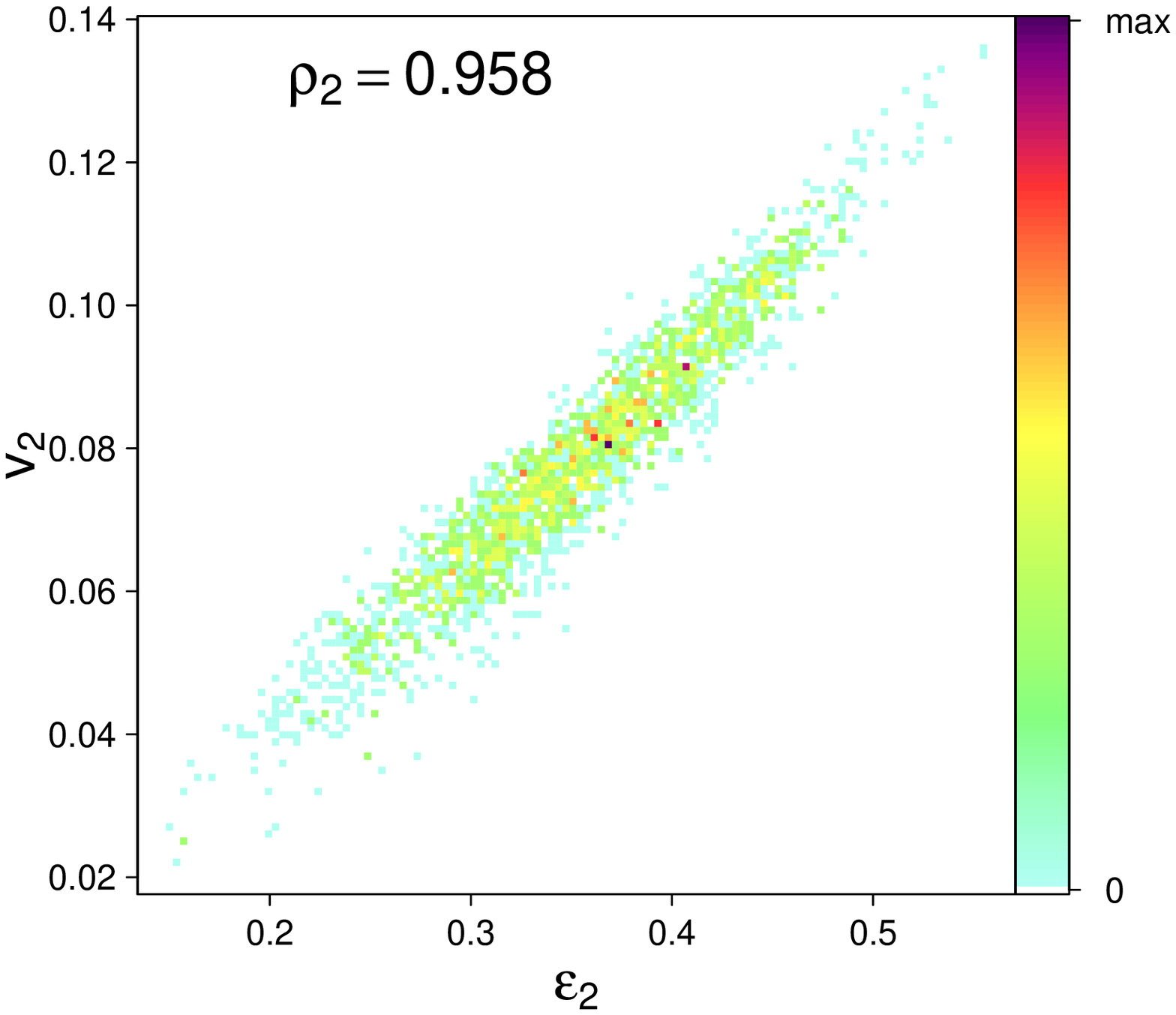}
\end{minipage}
\vspace{-20pt}
\\
\begin{minipage}{225pt}
\centering\includegraphics[width=220pt]{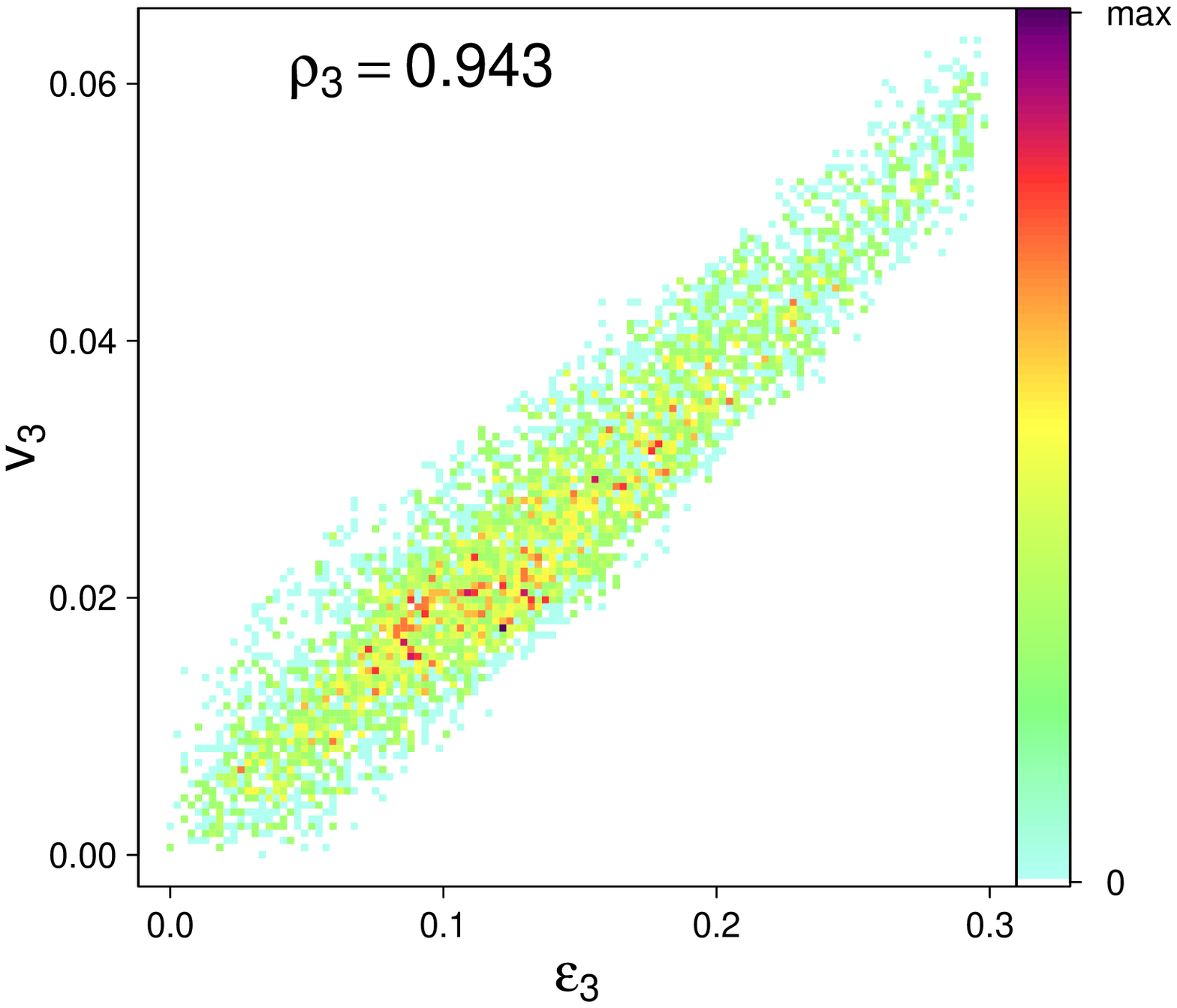}
\end{minipage}%
\begin{minipage}{225pt}
\centering\includegraphics[width=220pt]{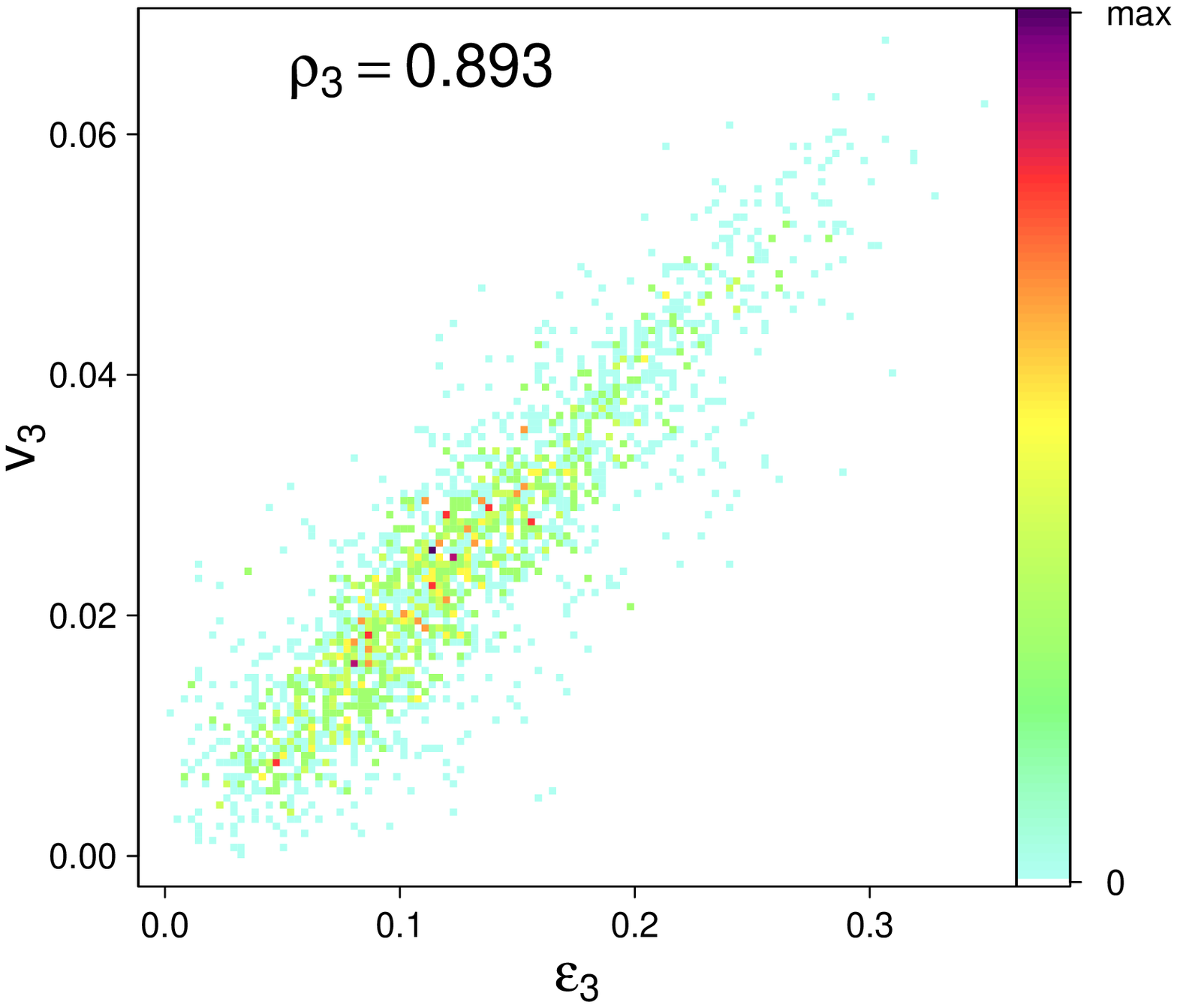}
\end{minipage}
\vspace{-20pt}
\\
\begin{minipage}{225pt}
\centering\includegraphics[width=220pt]{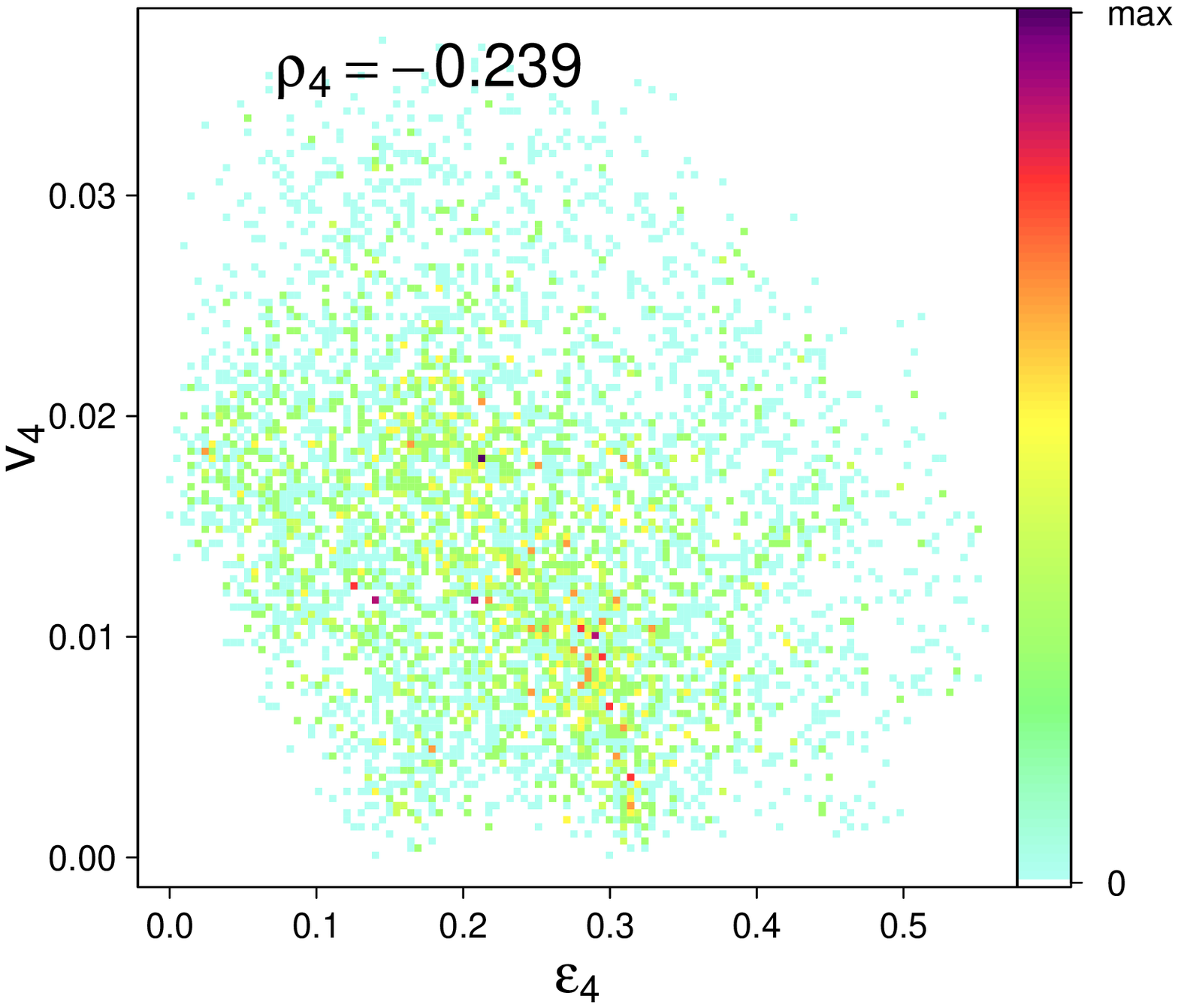}
\end{minipage}%
\begin{minipage}{225pt}
\centering\includegraphics[width=220pt]{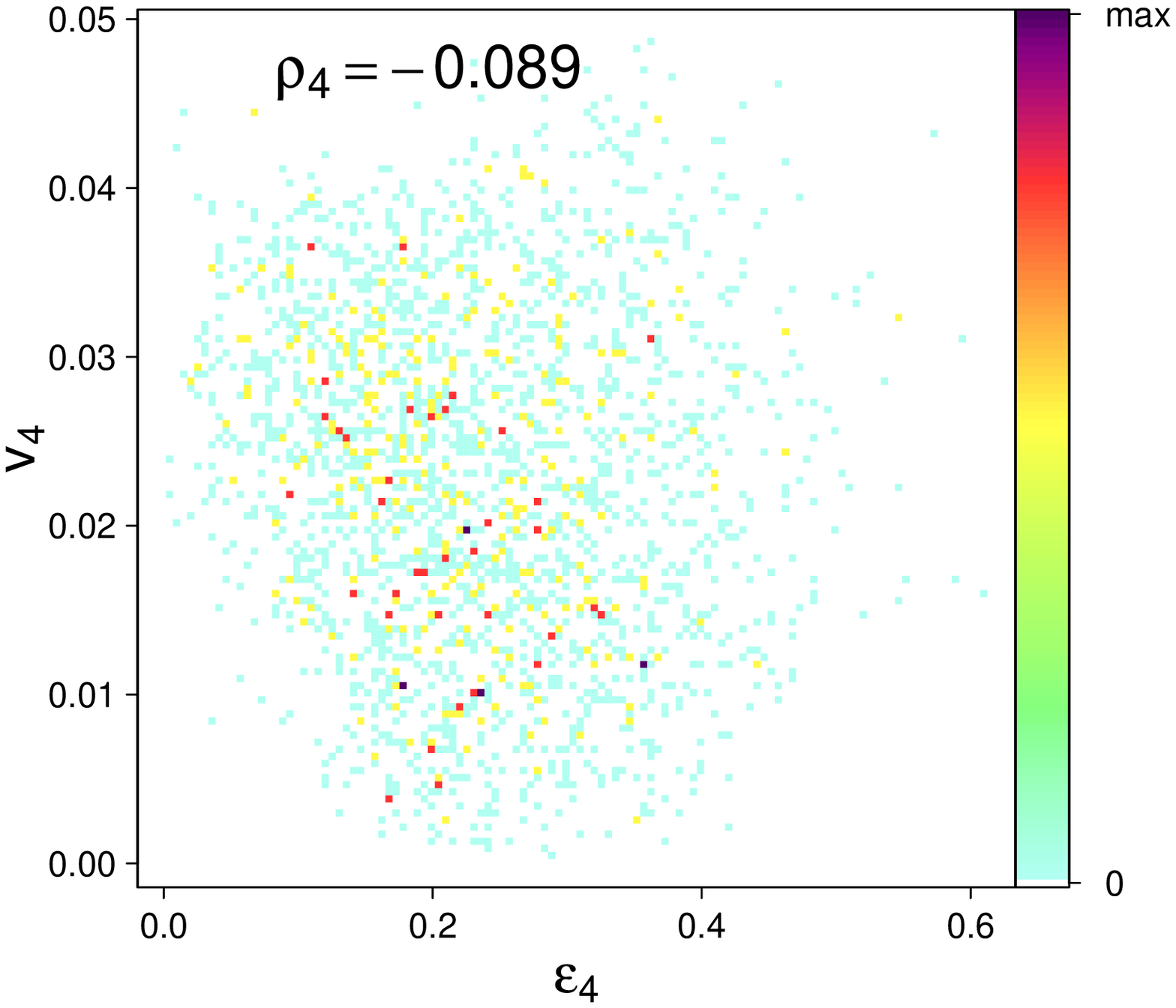}
\end{minipage}
\renewcommand{\figurename}{Fig.}
\caption{(Color online)
The calculated scatter plots of the event-by-event correlations between the flow harmonics $v_n$ and eccetricities $\varepsilon_n$.
The results are obtained by the peripheral tube model (left column) as well as by the anisotropic Gaussian model (right column).
In both cases, a total of 2000 events have been used to draw the plot.
The corresponding Pearson correlation coefficients are also given.}
\label{envn-tube-gaussj5}
\end{center}
\end{figure}

\begin{figure}[htbp]
\begin{center}
\begin{minipage}[c]{0.5\textwidth}
\centering\includegraphics[width=1.1\textwidth]{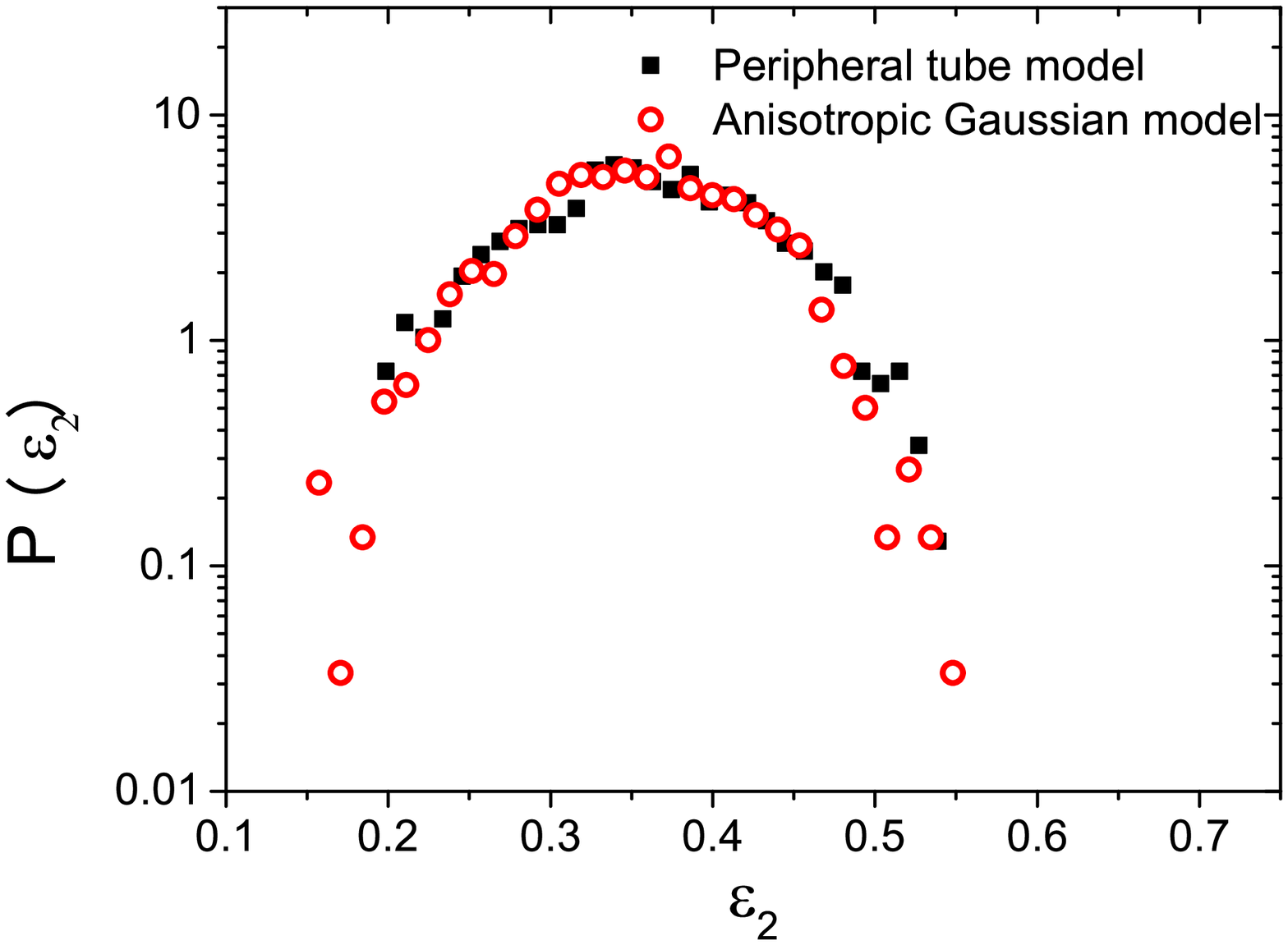}
\end{minipage}%
\begin{minipage}[c]{0.5\textwidth}
\centering\includegraphics[width=1.1\textwidth]{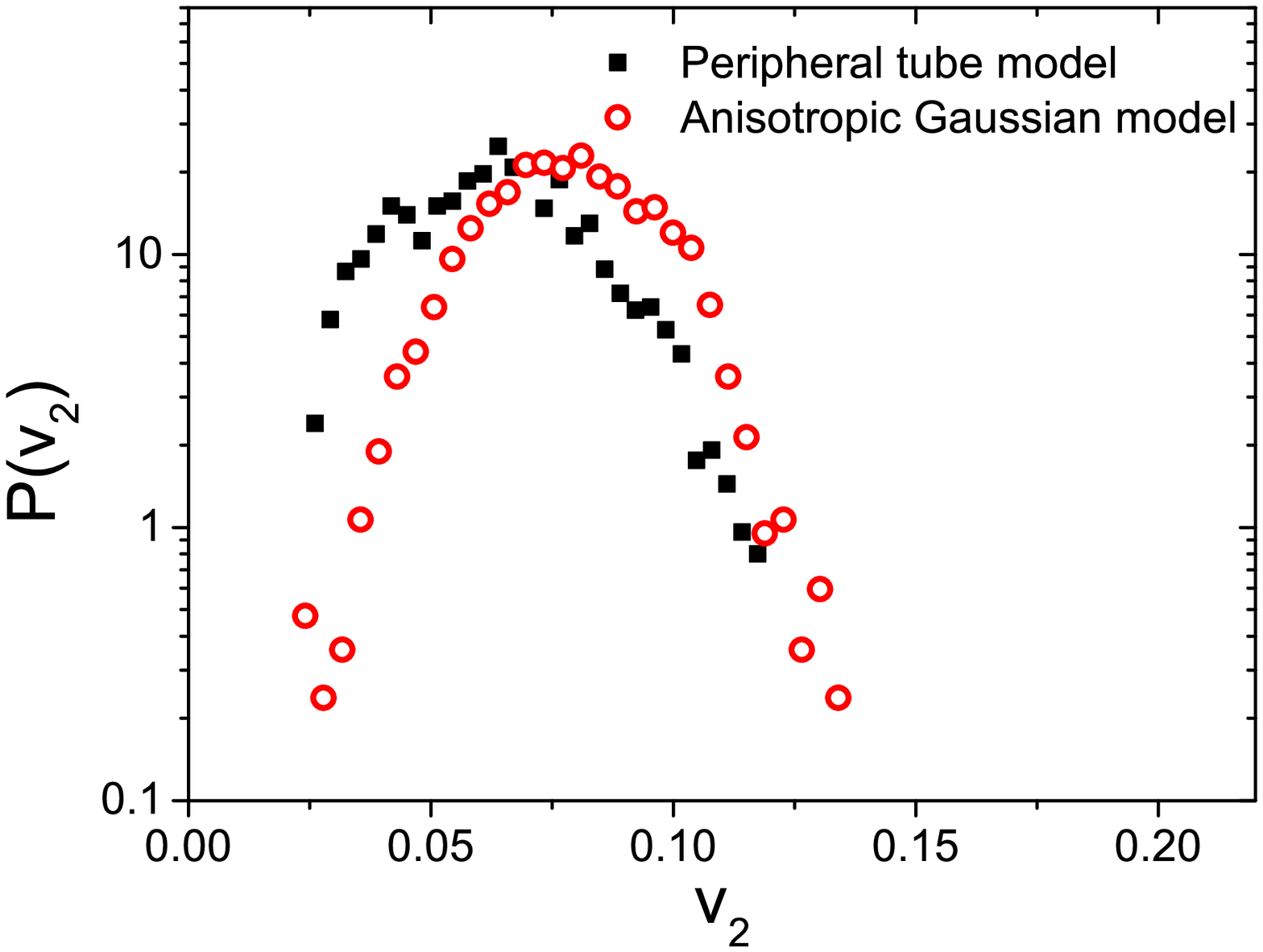}
\end{minipage}
\\
\begin{minipage}[c]{0.5\textwidth}
\centering\includegraphics[width=1.1\textwidth]{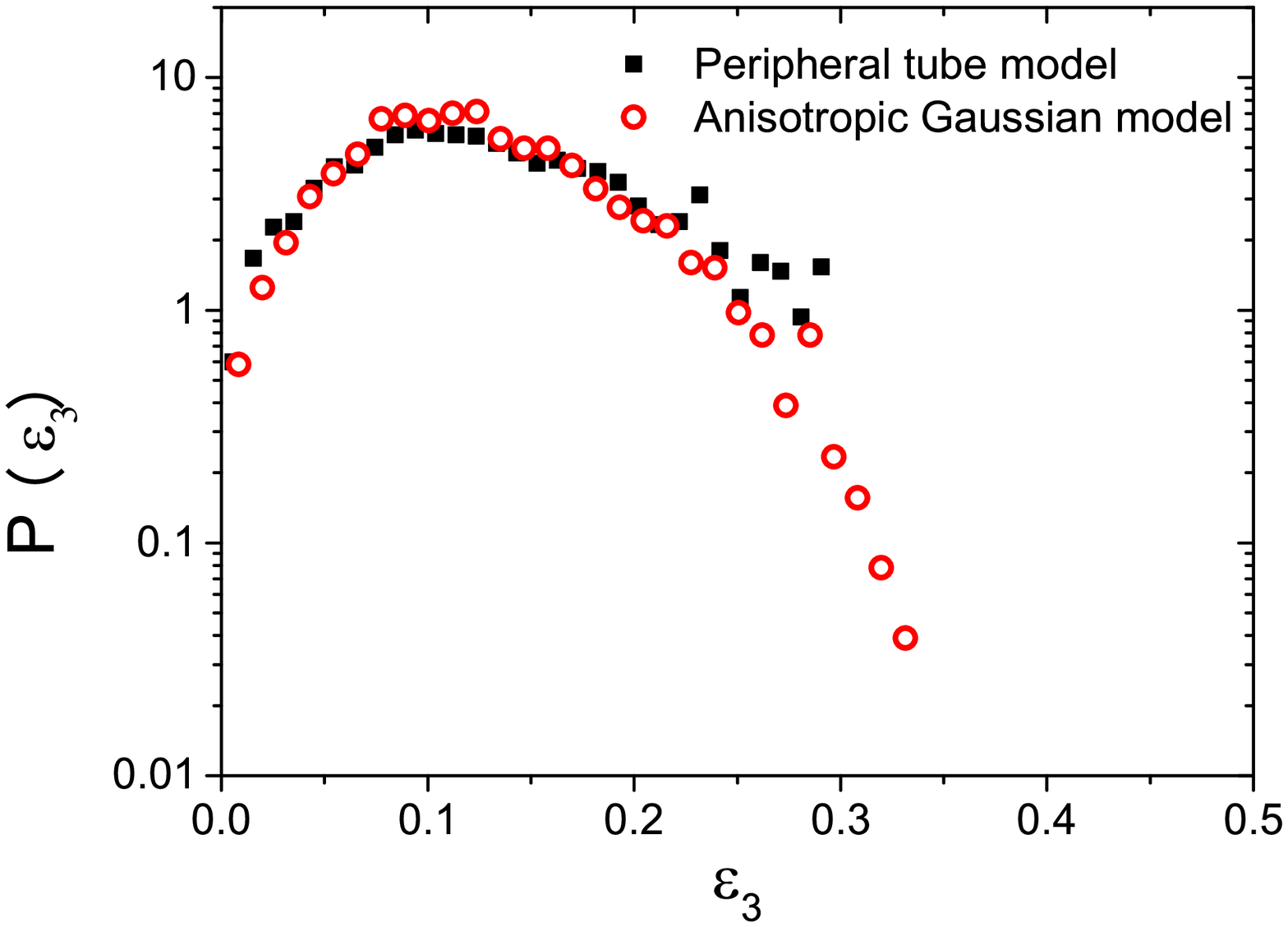}
\end{minipage}%
\begin{minipage}[c]{0.5\textwidth}
\centering\includegraphics[width=1.1\textwidth]{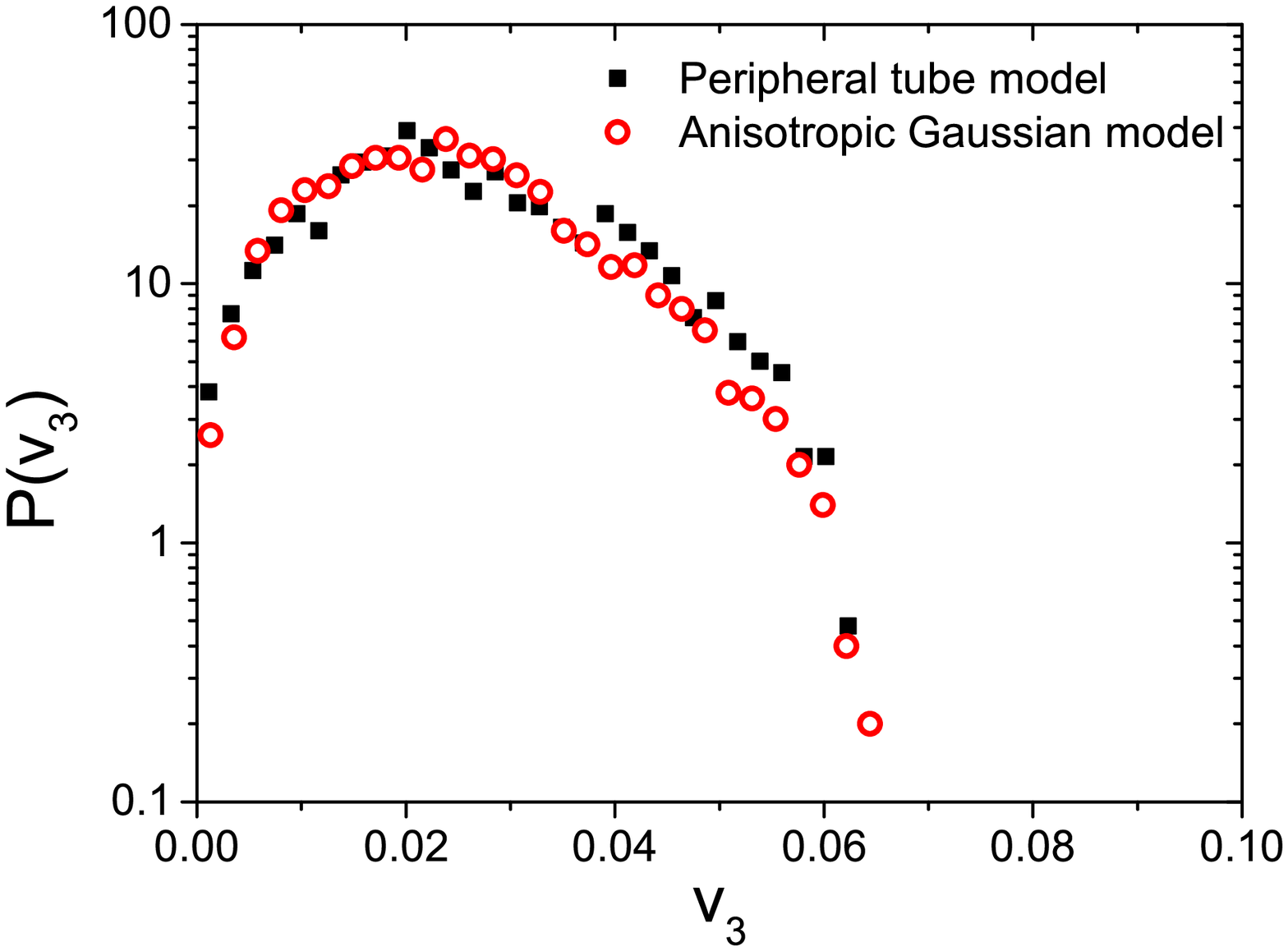}
\end{minipage}
\\
\begin{minipage}[c]{0.5\textwidth}
\centering\includegraphics[width=1.1\textwidth]{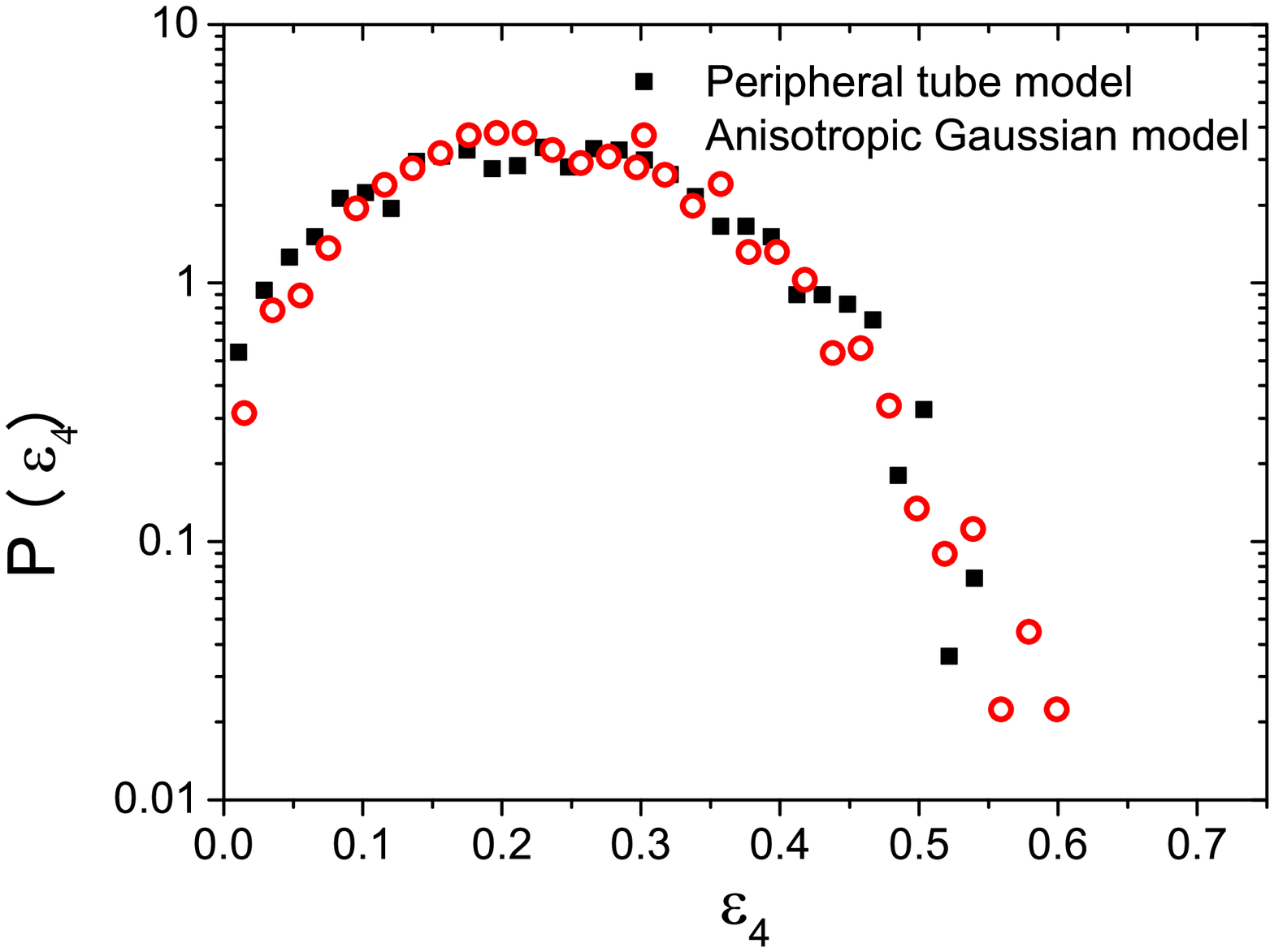}
\end{minipage}%
\begin{minipage}[c]{0.5\textwidth}
\centering\includegraphics[width=1.1\textwidth]{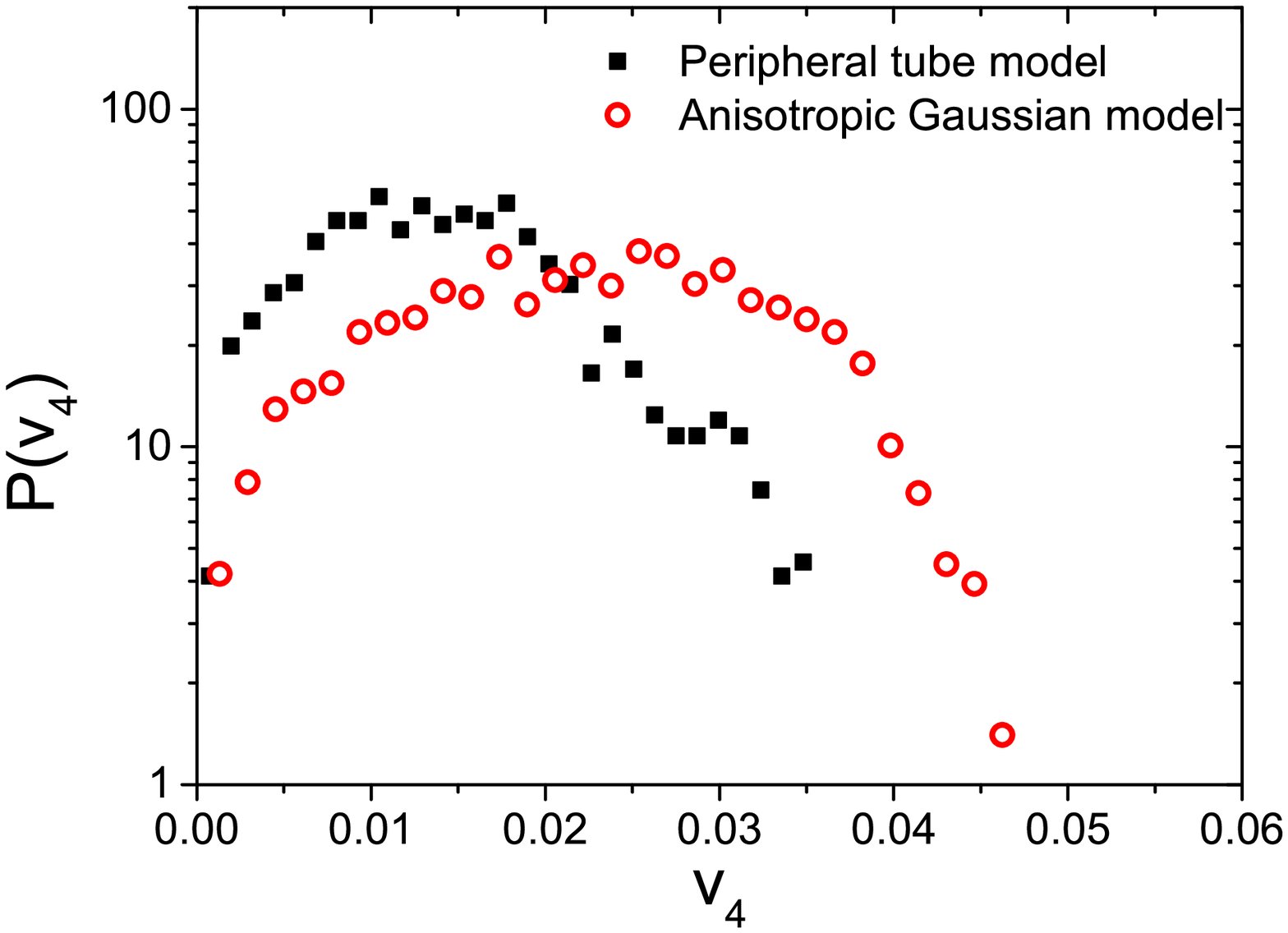}
\end{minipage}
\renewcommand{\figurename}{Fig.}
\caption{(Color online) 
Left: The probability density distribution of the event-by-event $\varepsilon_n$ in the peripheral tube model and anisotropic Gaussian model;
Right: The probability density distributions of the resultant flow harmonics $v_n$ in the two models.
The corresponding relative discrepancies are also evaluated.}
\label{penj5-pvnj5}
\end{center}
\end{figure}

The resultant relationship between eccentricities and flow harmonics for both models are shown in Fig.~\ref{envn-tube-gaussj5} and \ref{penj5-pvnj5}.
Fig.~\ref{envn-tube-gaussj5} presents the scatter plots regarding the relationship between the flow harmonics $v_n$ and eccetricities $\varepsilon_n$.
For each plot, we also evaluated the corresponding Pearson correlation coefficients defined by
\begin{eqnarray}
\rho_n\equiv \rho(v_n,\epsilon_n) = \frac{\mathrm{cov}(v_n,\epsilon_n)}{\sigma_{v_n}\sigma_{\epsilon_n}}  \, , \label{rhoLinearity} 
\end{eqnarray}
where $\mathrm{cov}(X,Y)$ and $\sigma_X$ are the covariance and standard deviations of the two quantities in question. 
As the resulting values of $\rho_n$ are bounded by $[-1, 1]$, larger value close to 1 implies a stronger positive linear correlation.
Fig.~\ref{envn-tube-gaussj5} indicates significant positive linear correlation in $\epsilon_2$ vs. $v_2$ and $\epsilon_3$ vs. $v_3$.
However, as observed in previsous studies~\cite{hydro-vn-04,hydro-vn-10}, the above correlations substantially decrease as $n$ further increases.
The magnitude of the resultant Pearson correlation coefficient drops nearly by an order of magnitude as one goes from $n=3$ to $n=4$.
A comparison between the tube model and the anisotropic Gaussian model shows that the linearities presented in the two models are mostly similar.
To be specific, the calculated Pearson correlation coefficients is slightly larger for the tube model for $n=2$ and $3$.
However, the difference becomes more significant in the case of $\epsilon_4$ vs. $v_4$.

To present the results from a different perspective, we show the probability density distributions of event-by-event eccentricities $\varepsilon_n$, as well as those of flow harmonics $v_n$, in Fig.~\ref{penj5-pvnj5}.
Here the calculated probability distributions from the two models are compared against each other. 
The plots in the left column of Fig.~\ref{penj5-pvnj5} give the probability density distributions of initial eccentricities.
This mostly serves to ensure quantitatively that the tuned models do possess ``similar" IC in terms of eccentricity components.
The right column, on the other hand, presents the resulting event-by-event distributions of flow harmonics.
Here, a sizable difference is observed in the case of the elliptic and quadrangular flow coefficients.
We note that this observation is not contradicting to the linearity that one may draw from Fig.~\ref{envn-tube-gaussj5}.
In fact, this is consistent with the previous findings, namely, the slope of the top-right plot of  Fig.~\ref{envn-tube-gaussj5} is slightly larger than that of the top-left plot.

\begin{figure}[htbp]
\begin{center}
\begin{minipage}[c]{0.5\textwidth}
\centering\includegraphics[width=1.1\textwidth]{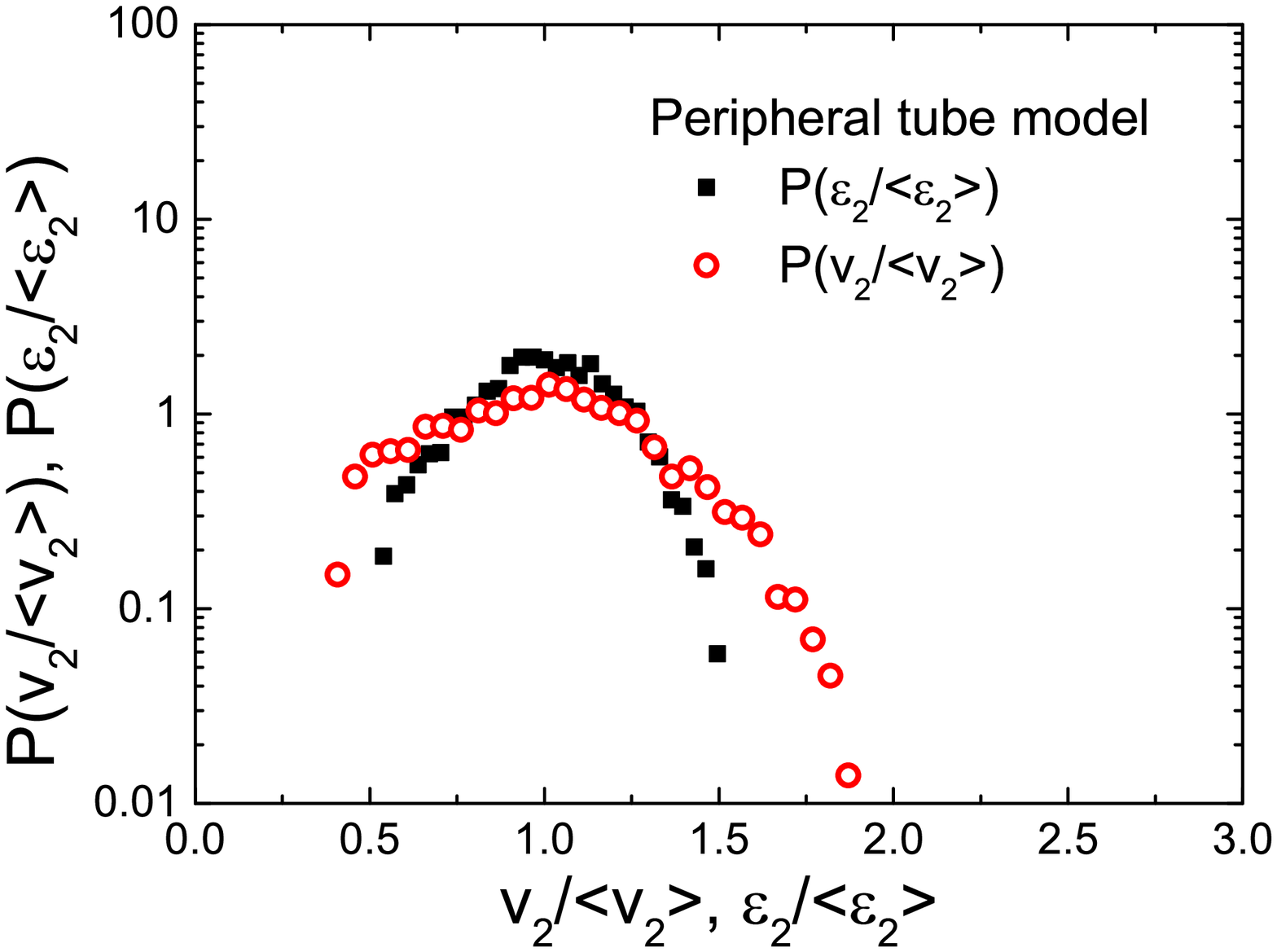}
\end{minipage}%
\begin{minipage}[c]{0.5\textwidth}
\centering\includegraphics[width=1.1\textwidth]{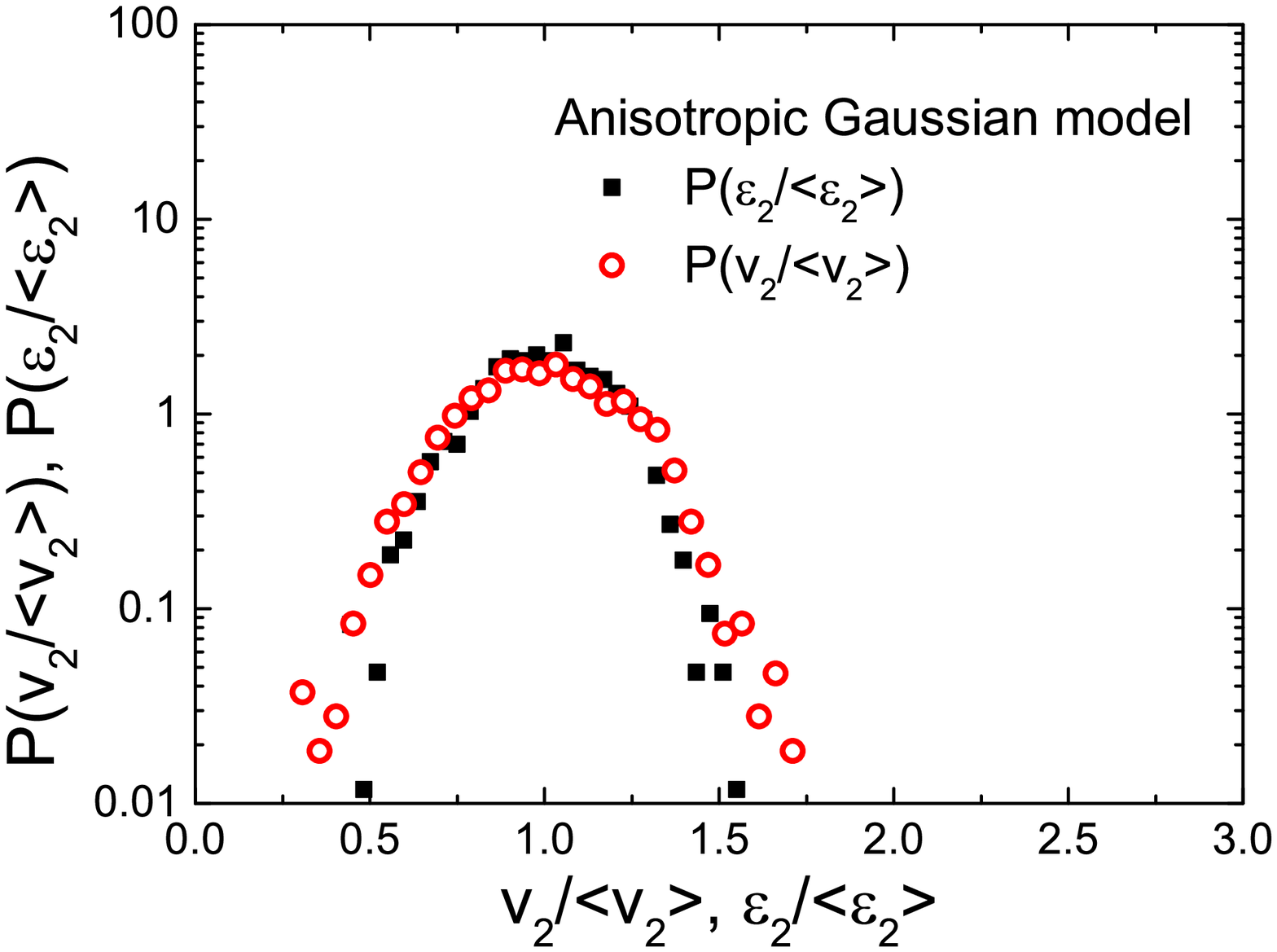}
\end{minipage}%
\\
\begin{minipage}[c]{0.5\textwidth}
\centering\includegraphics[width=1.1\textwidth]{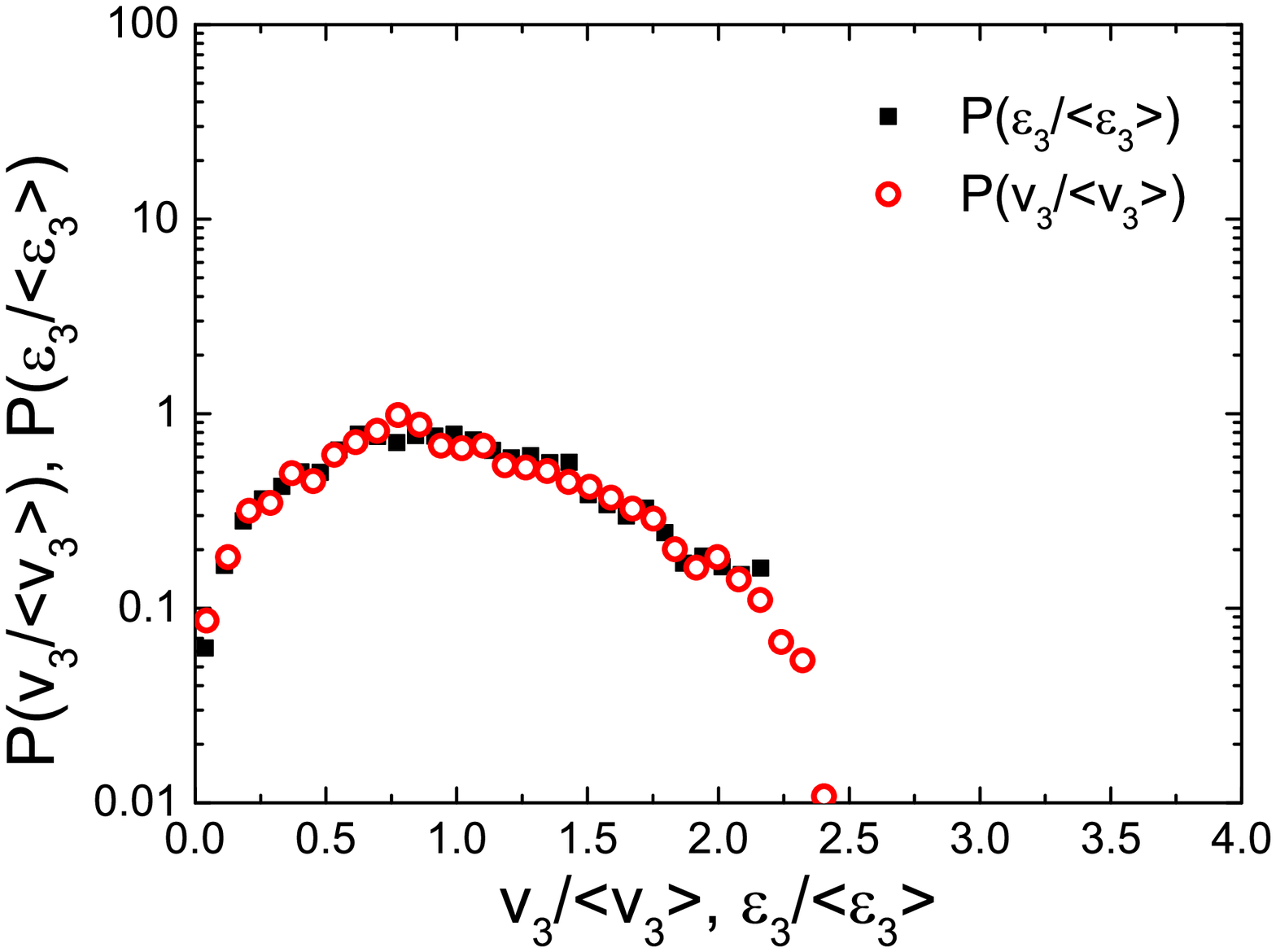}
\end{minipage}%
\begin{minipage}[c]{0.5\textwidth}
\centering\includegraphics[width=1.1\textwidth]{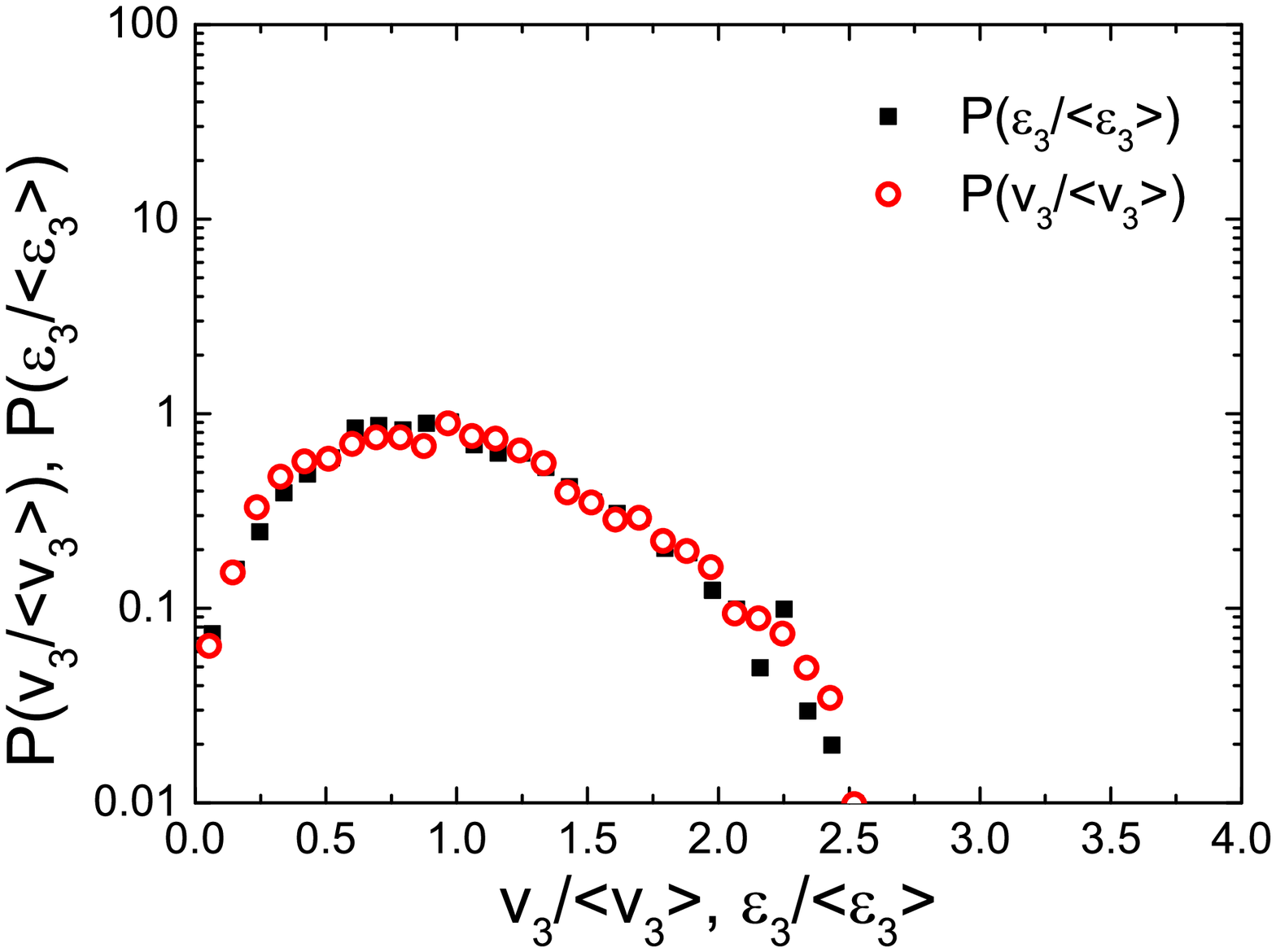}
\end{minipage}%
\\
\begin{minipage}[c]{0.5\textwidth}
\centering\includegraphics[width=1.1\textwidth]{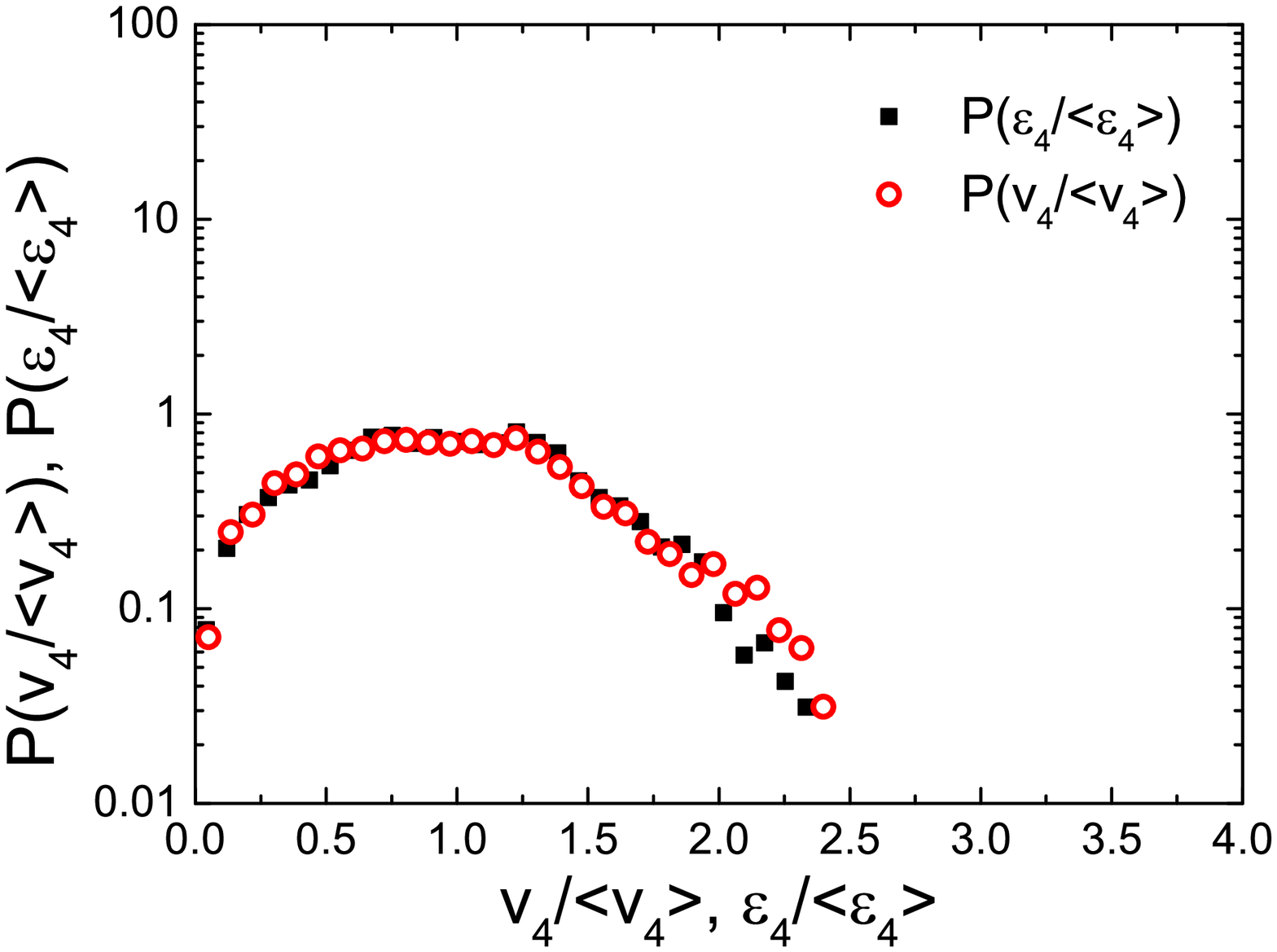}
\end{minipage}%
\begin{minipage}[c]{0.5\textwidth}
\centering\includegraphics[width=1.1\textwidth]{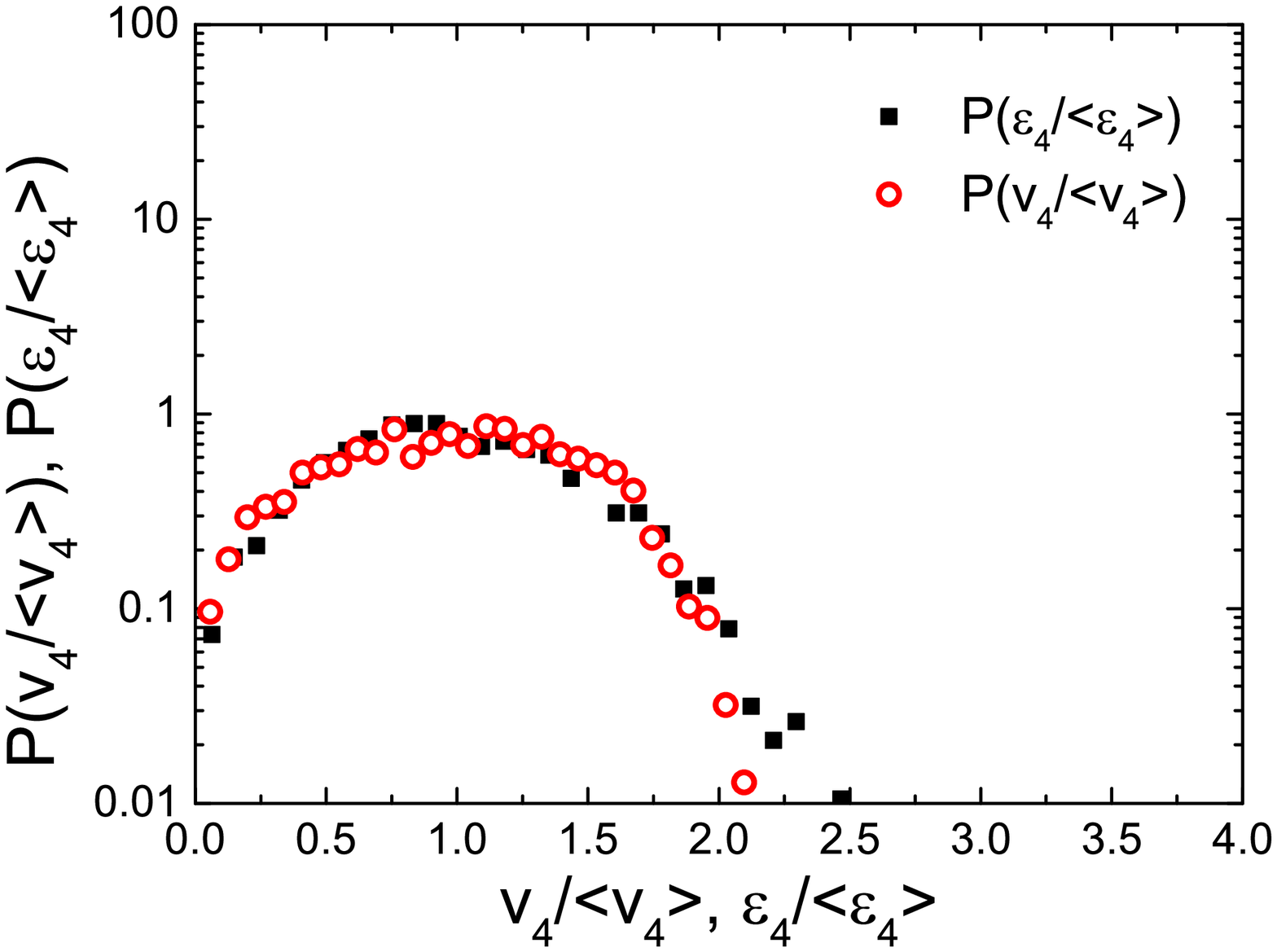}
\end{minipage}%
\renewcommand{\figurename}{Fig.}
\caption{(Color online)
The normalized probability density distribution of the event-by-event $\varepsilon_n$ and $v_n$ in the peripheral tube model (left column) and anisotropic Gaussian model (right column).}
\label{P_envn}
\end{center}
\end{figure}

To further investigate the linearity, we rescale the above results and present the normalized probability distributions in Fig.~\ref{P_envn}.
However, for the present purpose, we show in the same plot, the normalized distribution of $\epsilon_n$ against that of $v_n$.
In this case, we quantify the linearity by introducing the relative discrapancies of the two curves
\begin{eqnarray}
d_n\equiv d(P\left(v_n/\langle v_n\rangle\right),P\left(\epsilon_n/\langle \epsilon_n\rangle\right) = \frac{\int d\left(v_n/\langle v_n\rangle,\epsilon_n/\langle \epsilon_n\rangle\right)\frac{\left|P\left(v_n/\langle v_n\rangle\right)-P\left(\epsilon_n/\langle \epsilon_n\rangle\right)\right|}{\mathrm{max}\left(P\left(v_n/\langle v_n\rangle\right),P\left(\epsilon_n/\langle \epsilon_n\rangle\right)\right)}}{\int d\left(v_n/\langle v_n\rangle,\epsilon_n/\langle \epsilon_n\rangle\right)}  \, , \label{dnLinearity} 
\end{eqnarray}
which is evaluated for each plot.
It is found that linearity is mostly confirmed to a satisfactory degree, except for $d_2$ of the tube model.
We note that, in comparison with existing calculations~\cite{hydro-vn-05,hydro-vn-11,hydro-vn-10}, the peripheral tube model does present somewhat distinct features.
However, since eccentricity cannot be measured experimentally, the present findings do not contradict with any existing theory straightforwardly.
On the other hand, this result can be intuitively understood in terms of the peripheral tube model.
When a tube is located deep inside the system, the effect of its hydrodynamic expansion is mostly absorbed by the surrounding medium.
As a result, although it contributes significantly to the eccentricity, due to the smallness of its radial coordinate, it causes relatively insignificant impact to the flow harmonics.
On the contrary, a tube sitting close to the surface possesses the precisely opposite characteristic.
It leads to a significant disturbance to the one-particle distribution, resulting in sizable inhomogeneity in the media, while contributes little to the initial eccentricity.
As the IC configuration exaggerate the above feature, to some extent, its subsequent manifestation observed in Fig.~\ref{P_envn} is expected.

\begin{figure}[h]
\begin{center}
\begin{minipage}[c]{0.5\textwidth}
\centering\includegraphics[width=1.1\textwidth]{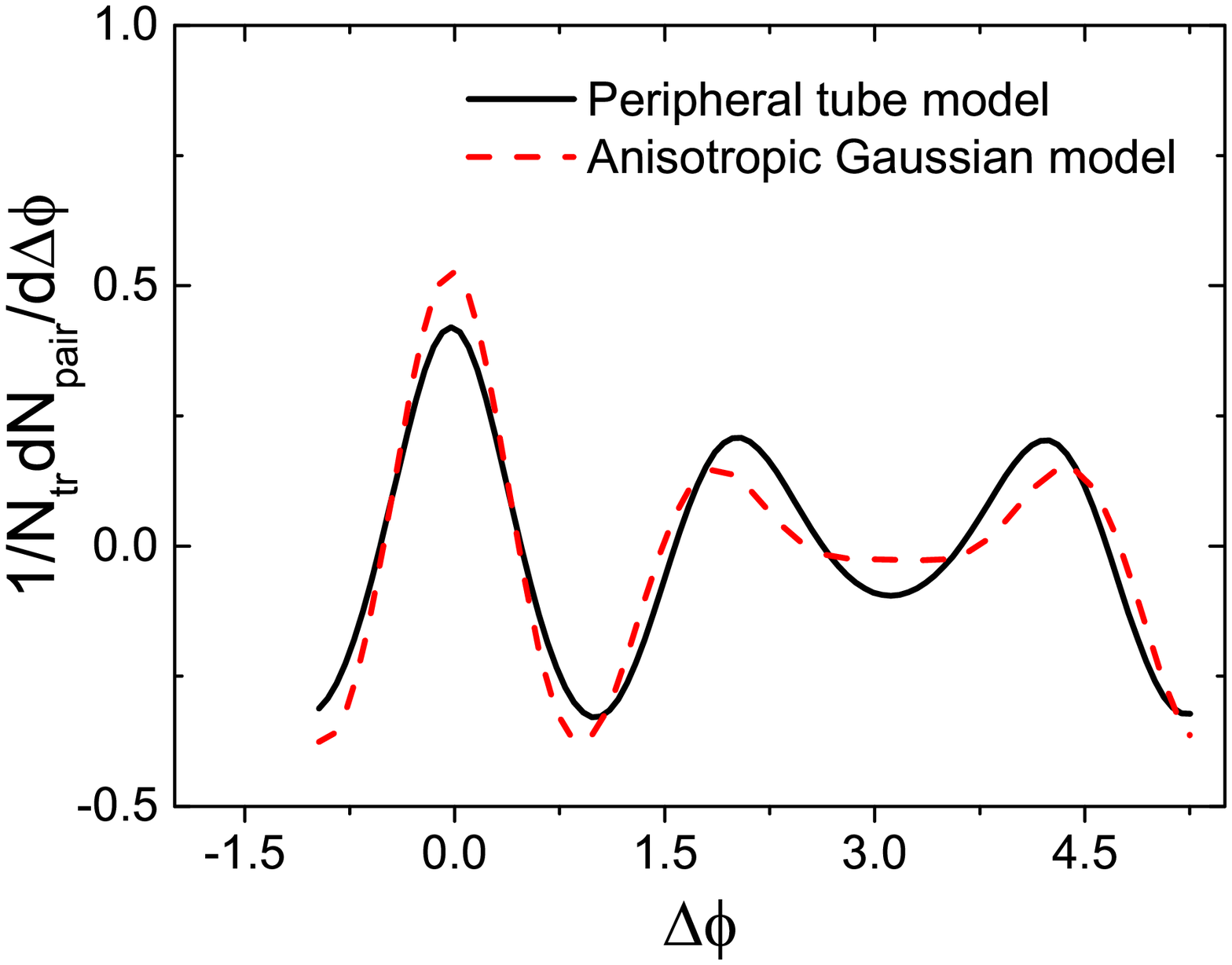}
\end{minipage}
\renewcommand{\figurename}{Fig.}
\caption{A comparison of the di-hardron correlations for $0.4<p_\mathrm{associated}<1 $ and $2<p_\mathrm{trigger}<3 $ for the peripheral tube model and anisotropic Gaussian model.}
\label{ridgepbpb}
\end{center}
\end{figure}

In Fig.~\ref{ridgepbpb}, we evaluate the di-hadron correlations for the peripheral tube model and the anisotropic Gaussian model.
We note that the resulting correlations shown in Fig.~\ref{ridgepbpb} do not attains zero at the minimum.
This is simply because the IC prepared in the present study does not contain any fluctuation in total entropy, and therefore, the resultant multiplicity fluctuations are minimized.
In fact, one can numerically check that the correlations presented in Fig.~\ref{ridgepbpb} integrate to zero over a period $0 < \Delta\phi \le 2\pi $.
As primarily determined by the average elliptic and triangular flows, the main features of the obtained di-hadron correlations are very similar among different models, consistent with previous studies~\cite{hydro-v3-01,hydro-v3-04,hydro-v3-05,hydro-v3-06,hydro-v3-07,hydro-v3-08,sph-corr-ev-10,sph-corr-ev-04,sph-corr-ev-06,sph-corr-ev-08}.

Now, we move to the study of linear and nonlinear response coefficients and other observables related to higher moments.
By making use of the complex anisotropic flow coefficient~\cite{hydro-corr-ph-08}
\begin{eqnarray}
P(\phi)=\frac{1}{2\pi}\sum^{+\infty}_{n=-\infty}V_n e^{-in\phi}
\end{eqnarray}
where $V_n=v_n \exp(in\Psi_n)$, and $v_n=|V_n|$.
Subsequently, one may study the nonlinear response coefficients given by~\cite{hydro-corr-ph-08,hydro-corr-04}
\begin{eqnarray}
\chi_4&=&\frac{\langle V_4(V^\ast_2)^2 \rangle}{\langle |V_2|^4 \rangle}=\frac{\langle v_4v_2^2\cos(4[\Psi_4-\Psi_2]) \rangle}{\langle v_2^4 \rangle},    \label{X4} \\
\chi_5&=&\frac{\langle V_5 V^\ast_2 V^\ast_3 \rangle}{\langle |V_2|^2 |V_3|^2 \rangle}=\frac{\langle v_5v_2v_3\cos(5\Psi_5-2\Psi_2-3\Psi_3) \rangle}{\langle v_2^2v_3^2 \rangle},     \label{X5} \\
\chi_{62}&=&\frac{\langle V_6(V^\ast_2)^3 \rangle}{\langle |V_2|^6 \rangle}=\frac{\langle v_6v_2^3\cos(6[\Psi_6-\Psi_2]) \rangle}{\langle v_2^6 \rangle},     \label{X62}\\
\chi_{63}&=&\frac{\langle V_6(V^\ast_3)^2 \rangle}{\langle |V_3|^4 \rangle}=\frac{\langle v_6v_3^2\cos(6[\Psi_6-\Psi_3]) \rangle}{\langle v_3^4 \rangle},     \label{X63}\\
\chi_7&=&\frac{\langle V_7(V^\ast_2)^2 V^\ast_3 \rangle}{\langle |V_2|^4 |V_3|^2 \rangle}=\frac{\langle v_7v_2^2v_3\cos(7\Psi_7-4\Psi_2-3\Psi_3) \rangle}{\langle v_2^4v_3^2 \rangle},     \label{X7}
\end{eqnarray}
where, for instance, the imaginal part of the first expression $\langle v_4v_2^2\sin(4[\Psi_4-\Psi_2]) \rangle = 0$ for a large number of events.
In Tab.~\ref{nonlinearcoeffpbpb}, we present the calculated nonlinear response coefficients evaluated for the two models in comparison to those extracted from CMS and ATLAS data~\cite{hydro-corr-05,LHC-cms-vn-3,LHC-atlas-vn-3}.
Though the results are mostly of the same order of magnitude when compared to the experimental values, the discrepancies are attributed to the fact that the IC considered in the present study are not realistic.
The results between the peripheral model and anisotropic Gaussian model are, on the other hand, obtained by IC with essentially identical probability distribution of eccentricity.
In this context, although the difference between the two models are of similar magnitude, one observes that the difference is sizable.
Comparing to the linear response presented above, we note that the difference between the two models regarding nonlinear response coefficients is more substantial.

\begin{table}[h!]
\begin{center}
  \caption{The calculated nonlinear response coefficients for $2.76$ TeV Pb+Pb collisions for the 20\%-25\% centrality class.}
  \begin{tabular}{c|c|c|c|c|c}
  \hline
  &  Number of events    & $\chi_4$    &    $\chi_5$   &    $\chi_{62}$    &    $\chi_{63}$            \\
  \hline
  \multirow{1}{*}{\tabincell{c}{CMS and ATLAS data}}
  &                      & 0.818 &  1.878     &   0.715          &  0.878                 \\
  \hline
  \multirow{1}{*}{\tabincell{c}{Peripheral tube model }}
  &   (2000 events)      & 1.50 &  2.44     &   3.46          &  2.33               \\
  \hline
  \multirow{1}{*}{\tabincell{c}{Anisotropic Gaussian model}}
  &   (2200 events)      & 1.69 &  2.794     &   2.97          &  1.12                \\
  \hline
  \end{tabular}
  \label{nonlinearcoeffpbpb}
\end{center}
\end{table}

In Tab.~\ref{symmetric-cumulant}, we present the results on symmetric cumulants as well as normalized symmetric cumulants, calculated by using the following definitions.
\begin{eqnarray}
SC(m,n)=\langle v^2_nv^2_m \rangle-\langle v^2_n \rangle\langle v^2_m  \rangle, \\
NSC(m,n)=\frac{\langle v^2_nv^2_m \rangle-\langle v^2_n \rangle\langle v^2_m  \rangle}{\langle v^2_n \rangle\langle  v^2_m \rangle} .
\end{eqnarray}
The normalized symmetric cumulant is understood as a measure for the correlation of the magnitude of flow fluctuations.
The mixed harmonics, on the other hand, are related to the ratios between flow harmonics evaluated by different event planes.
Lastly, we show in Tab.~\ref{mixed-harmonics} the calcualted mixed cumulants.
It is found that the difference between the two models is significant.
Especially for $SC(4,2)$ and $NSC(4,2)$, the signs of the correlations are opposite for the two models.
We understand that the observed difference residing in higher harmonics and nonlinear response coefficients are originated from the distinction between these two models.

\begin{table}[htbp]
\begin{center}
  \caption{The calculated symmetric cumulants for $2.76$ TeV Pb+Pb collisions in $20\%-25\%$ centrality class.}
  \begin{tabular}{c|c|c|c|c}
  \hline
       & $SC(4,2)$    &    $SC(3,2)$   &    $NSC(4,2)$    &    $NSC(3,2)$              \\
  \hline
  \multirow{1}{*}{\tabincell{c}{Peripheral tube model}}
       & $0.043\times 10^{-6}$ & $ 0.134\times 10^{-6}$     &   0.035          &  0.035               \\
  \hline
  \multirow{1}{*}{\tabincell{c}{Anisotropic Gaussian model}}
       & $-0.252\times 10^{-6}$ &  $0.0552\times 10^{-6} $    &    -0.063         &  0.011               \\
  \hline
  \end{tabular}
  \label{symmetric-cumulant}
\end{center}
\end{table}

\begin{table}[htbp]
\begin{center}
  \caption{The mixed harmonics for $2.76$ TeV Pb+Pb collisions in $20\%-25\%$ centrality class.}
  \begin{tabular}{c|c|c}
  \hline
      &   Peripheral tube model    &    Anisotropic Gaussian model                       \\
  \hline
  \multirow{1}{*}{\tabincell{c}{$\langle  v_4v_2^2\cos(4[\Psi_4-\Psi_2]) \rangle$}}
      &   $4.09\times 10^{-5}$     &   $8.39\times 10^{-5}$               \\
  \hline
  \multirow{1}{*}{\tabincell{c}{$\langle  v_5v_2v_3\cos(5\Psi_5-2\Psi_2-3\Psi_3) \rangle$}}
      &   $0.954\times 10^{-5}$    &   $1.38\times 10^{-5}$                \\
  \hline
  \multirow{1}{*}{\tabincell{c}{$\langle  v_6v_2^3\cos(6[\Psi_6-\Psi_2]) \rangle$}}
      &  $0.0721\times 10^{-5}$    &    $ 0.129\times 10^{-5}$               \\
  \hline
   \multirow{1}{*}{\tabincell{c}{$\langle  v_6v_3^2\cos(6[\Psi_6-\Psi_3]) \rangle$}}
      &   $0.301\times 10^{-5}$    &    $0.119\times 10^{-5}$              \\
  \hline
  \end{tabular}
  \label{mixed-harmonics}
\end{center}
\end{table}

\section{IV. Concluding remarks}

To summarize, in this work, we devised an anisotropic Gaussian model to match the eccentricity probability distribution of the peripheral tube model.
By doing this, we carried out a back-to-back comparison between the two models regarding the mapping between the event-by-event IC fluctuations and flow harmonics.
In particular, we studied the linear as well as the nonlinear response of the system in terms of flow harmonic coefficients, di-hadron correlations, symmetric cumulants, mixed harmonics, among others.
Although the di-hadron correlations seem similar in their shapes, the distinction between the two models can be revealed by more detailed observables.
In particular, the discrepancies in the normalized probability distributions of $\epsilon_2$ and $v_2$ can be readily understood in terms of the physical nature of the peripheral tube model.
Furthermore, the calculated Pearson correlation coefficient regarding higher-order harmonics also demonstrated a substantial difference between the two models.
In this context, it might be interesting to follow this train of thought by proposing observables, which may quantify the nonlinearity to a greater extent.

From a hydrodynamic point of view, one has to deal with physical concepts such as the degree of local thermal equilibrium, the equation of state, and transport coefficients.
Subsequently, one must choose appropriate tools that reflect the characteristics of the long-wavelength limit of a strongly interacting system, which is, by and large, genuinely nonlinear.  
In this context, the subtle difference which carries the vital information may reside in quantities such as higher-order correlators~\cite{hydro-corr-ph-07,hydro-corr-ph-08,hydro-vn-10}.
In particular, the deviation from the linearity, even though it can be insignificant in magnitude, might be particularly meaningful
This is because, in the framework of the event-by-event fluctuations, a state close to the local thermal equilibrium only corresponds to a tiny space-time domain during the entire dynamical evolution (Ref.~\cite{hydro-review-06,phsd-01}). 
This topic is also closely related to the question concerning to what degree the genuine event-by-event hydrodynamics is feasible.
Besides the study of the deviations from linearity, one may also look for quantities that are intrinsically associated with the nonlinear nature of the system.
Further studies concerning this topic are in progress.

\section*{Acknowledgments}
We gratefully acknowledge the financial support from
Funda\c{c}\~ao de Amparo \`a Pesquisa do Estado de S\~ao Paulo (FAPESP),
Funda\c{c}\~ao de Amparo \`a Pesquisa do Estado do Rio de Janeiro (FAPERJ),
Conselho Nacional de Desenvolvimento Cient\'{\i}fico e Tecnol\'ogico (CNPq),
and Coordena\c{c}\~ao de Aperfei\c{c}oamento de Pessoal de N\'ivel Superior (CAPES).
A part of the work was developed under the project INCT-FNA Proc. No. 464898/2014-5, the Center for Scientific Computing (NCC/GridUNESP) of the S\~ao Paulo State University (UNESP),
also, the National Natural Science Foundation of China (NNSFC) under contract No.11805166.

\bibliographystyle{h-physrev-qian}
\bibliography{references_qian}

\begin{thebibliography}{10}

\bibitem{hydro-review-04}
U.~W. Heinz and R.~Snellings,
\newblock Annu. Rev. Nucl. Part. Sci. {\bf 63}, 123 (2013), arXiv:1301.2826.

\bibitem{hydro-review-05}
C.~Gale, S.~Jeon, and B.~Schenke,
\newblock Int. J. Mod. Phys. {\bf A28}, 1340011 (2013), arXiv:1301.5893.

\bibitem{hydro-review-06}
R.~Derradi~de Souza, T.~Koide, and T.~Kodama,
\newblock Prog. Part. Nucl. Phys. {\bf 86}, 35 (2016), arXiv:1506.03863.

\bibitem{hydro-review-08}
T.~Hirano, P.~Huovinen, K.~Murase, and Y.~Nara,
\newblock Prog. Part. Nucl. Phys. {\bf 70}, 108 (2013), arXiv:1204.5814.

\bibitem{hydro-review-09}
T.~Kodama, H.~Stocker, and N.~Xu,
\newblock Journal of Physics G: Nuclear and Particle Physics {\bf 41}, 120301
  (2014).

\bibitem{hydro-review-10}
W.~Florkowski, M.~P. Heller, and M.~Spalinski,
\newblock Rept. Prog. Phys. {\bf 81}, 046001 (2018), arXiv:1707.02282.

\bibitem{hydro-v3-02}
D.~Teaney and L.~Yan,
\newblock Phys.Rev. {\bf C83}, 064904 (2011), arXiv:1010.1876.

\bibitem{hydro-vn-03}
D.~Teaney and L.~Yan,
\newblock Phys.Rev. {\bf C86}, 044908 (2012), arXiv:1206.1905.

\bibitem{sph-vn-03}
F.~G. Gardim, F.~Grassi, M.~Luzum, and J.-Y. Ollitrault,
\newblock Phys. Rev. {\bf C85}, 024908 (2012), arXiv:1111.6538.

\bibitem{hydro-vn-04}
H.~Niemi, G.~Denicol, H.~Holopainen, and P.~Huovinen,
\newblock Phys.Rev. {\bf C87}, 054901 (2012), arXiv:1212.1008.

\bibitem{sph-vn-04}
W.-L. Qian {\em et~al.},
\newblock J.Phys.G {\bf G41}, 015103 (2014), arXiv:1305.4673.

\bibitem{sph-vn-06}
F.~G. Gardim, F.~Grassi, P.~Ishida, M.~Luzum, and J.-Y. Ollitrault,
\newblock Phys. Rev. {\bf C100}, 054905 (2019), arXiv:1906.03045.

\bibitem{hydro-vn-10}
J.~Fu,
\newblock Phys. Rev. {\bf C92}, 024904 (2015).

\bibitem{hydro-vn-ph-09}
S.~Floerchinger and U.~A. Wiedemann,
\newblock Phys. Rev. {\bf C88}, 044906 (2013), arXiv:1307.7611.

\bibitem{hydro-vn-ph-10}
C.~E. Coleman-Smith, H.~Petersen, and R.~L. Wolpert,
\newblock J. Phys. {\bf G40}, 095103 (2013), arXiv:1204.5774.

\bibitem{hydro-vn-ph-11}
S.~Floerchinger and U.~A. Wiedemann,
\newblock Phys. Lett. {\bf B728}, 407 (2014), arXiv:1307.3453.

\bibitem{hydro-corr-ph-06}
A.~Bilandzic, C.~H. Christensen, K.~Gulbrandsen, A.~Hansen, and Y.~Zhou,
\newblock Phys. Rev. {\bf C89}, 064904 (2014), arXiv:1312.3572.

\bibitem{hydro-corr-ph-07}
R.~S. Bhalerao, J.-Y. Ollitrault, and S.~Pal,
\newblock Phys. Rev. {\bf C88}, 024909 (2013), arXiv:1307.0980.

\bibitem{hydro-corr-ph-08}
R.~S. Bhalerao, J.-Y. Ollitrault, and S.~Pal,
\newblock Phys. Lett. {\bf B742}, 94 (2015), arXiv:1411.5160.

\bibitem{hydro-corr-04}
L.~Yan and J.-Y. Ollitrault,
\newblock Phys. Lett. {\bf B744}, 82 (2015), arXiv:1502.02502.

\bibitem{hydro-corr-05}
L.~Yan, S.~Pal, and J.-Y. Ollitrault,
\newblock Nucl. Phys. {\bf A956}, 340 (2016), arXiv:1601.00040.

\bibitem{LHC-cms-vn-3}
CMS, S.~Chatrchyan {\em et~al.},
\newblock Phys. Rev. {\bf C89}, 044906 (2014), arXiv:1310.8651.

\bibitem{LHC-atlas-vn-3}
ATLAS, G.~Aad {\em et~al.},
\newblock Phys. Rev. {\bf C90}, 024905 (2014), arXiv:1403.0489.

\bibitem{zml-pca-01}
J.~Shlens,
\newblock CoRR {\bf abs/1404.1100} (2014), arXiv:1404.1100.

\bibitem{hydro-vn-pca-01}
R.~S. Bhalerao, J.-Y. Ollitrault, S.~Pal, and D.~Teaney,
\newblock Phys. Rev. Lett. {\bf 114}, 152301 (2015), arXiv:1410.7739.

\bibitem{hydro-vn-pca-02}
P.~Bozek,
\newblock Phys. Rev. {\bf C97}, 034905 (2018), arXiv:1711.07773.

\bibitem{hydro-vn-pca-03}
M.~Hippert {\em et~al.},
\newblock Phys. Rev. {\bf C101}, 034903 (2020), arXiv:1906.08915.

\bibitem{hydro-vn-pca-04}
Z.~Liu, A.~Behera, H.~Song, and J.~Jia,
\newblock (2020), arXiv:2002.06061.

\bibitem{sph-corr-02}
Y.~Hama, R.~P.~G. Andrade, F.~Grassi, and W.-L. Qian,
\newblock Nonlin.Phenom.Complex Syst. {\bf 12}, 466 (2009), arXiv:0911.0811.

\bibitem{sph-corr-03}
R.~Andrade, F.~Grassi, Y.~Hama, and W.-L. Qian,
\newblock J.Phys.G {\bf G37}, 094043 (2010), arXiv:0912.0703.

\bibitem{sph-corr-04}
R.~P.~G. Andrade, F.~Grassi, Y.~Hama, and W.-L. Qian,
\newblock Phys.Lett. {\bf B712}, 226 (2012), arXiv:1008.4612.

\bibitem{sph-corr-07}
Y.~Hama, R.~P. Andrade, F.~Grassi, J.~Noronha, and W.-L. Qian,
\newblock Acta Phys.Polon.Supp. {\bf 6}, 513 (2013), arXiv:1212.6554.

\bibitem{sph-corr-08}
D.~Wen {\em et~al.},
\newblock J. Phys. {\bf G46}, 035103 (2019), arXiv:1808.03775.

\bibitem{nexus-1}
H.~Drescher, S.~Ostapchenko, T.~Pierog, and K.~Werner,
\newblock Phys.Rev. {\bf C65}, 054902 (2002), arXiv:hep-ph/0011219.

\bibitem{nexus-rept}
H.~Drescher, M.~Hladik, S.~Ostapchenko, T.~Pierog, and K.~Werner,
\newblock Phys.Rept. {\bf 350}, 93 (2001), arXiv:hep-ph/0007198.

\bibitem{epos-1}
K.~Werner, F.-M. Liu, and T.~Pierog,
\newblock Phys.Rev. {\bf C74}, 044902 (2006), arXiv:hep-ph/0506232.

\bibitem{epos-2}
K.~Werner, I.~Karpenko, and T.~Pierog,
\newblock Phys.Rev.Lett. {\bf 106}, 122004 (2011), arXiv:1011.0375.

\bibitem{epos-3}
K.~Werner, M.~Bleicher, B.~Guiot, I.~Karpenko, and T.~Pierog,
\newblock Phys. Rev. Lett. {\bf 112}, 232301 (2014), arXiv:1307.4379.

\bibitem{hydro-vn-05}
C.~Gale, S.~Jeon, B.~Schenke, P.~Tribedy, and R.~Venugopalan,
\newblock Phys.Rev.Lett. {\bf 110}, 012302 (2013), arXiv:1209.6330.

\bibitem{hydro-vn-11}
P.~Bozek and W.~Broniowski,
\newblock Phys. Rev. {\bf C88}, 014903 (2013), arXiv:1304.3044.

\bibitem{hydro-v3-01}
B.~Alver and G.~Roland,
\newblock Phys.Rev. {\bf C81}, 054905 (2010), arXiv:1003.0194.

\bibitem{hydro-v3-04}
G.-Y. Qin, H.~Petersen, S.~A. Bass, and B.~Muller,
\newblock Phys.Rev. {\bf C82}, 064903 (2010), arXiv:1009.1847.

\bibitem{hydro-v3-05}
B.~Schenke, S.~Jeon, and C.~Gale,
\newblock Phys.Rev.Lett. {\bf 106}, 042301 (2011), arXiv:1009.3244.

\bibitem{hydro-v3-06}
J.~Xu and C.~M. Ko,
\newblock Phys.Rev. {\bf C83}, 021903 (2011), arXiv:1011.3750.

\bibitem{hydro-v3-07}
G.-L. Ma and X.-N. Wang,
\newblock Phys.Rev.Lett. {\bf 106}, 162301 (2011), arXiv:1011.5249.

\bibitem{hydro-v3-08}
H.~Petersen, G.-Y. Qin, S.~A. Bass, and B.~Muller,
\newblock Phys. Rev. {\bf C82}, 041901 (2010), arXiv:1008.0625.

\bibitem{sph-corr-ev-10}
R.~P.~G. Andrade, F.~Grassi, Y.~Hama, and W.~L. Qian,
\newblock Nucl. Phys. {\bf A854}, 81 (2011), arXiv:1008.0139.

\bibitem{sph-corr-ev-04}
W.-L. Qian, R.~Andrade, F.~Gardim, F.~Grassi, and Y.~Hama,
\newblock Phys.Rev. {\bf C87}, 014904 (2013), arXiv:1207.6415.

\bibitem{sph-corr-ev-06}
W.~M. Castilho, W.-L. Qian, F.~G. Gardim, Y.~Hama, and T.~Kodama,
\newblock Phys.Rev. {\bf C95}, 064908 (2017), arXiv:1610.04108.

\bibitem{sph-corr-ev-08}
W.~M. Castilho, W.-L. Qian, Y.~Hama, and T.~Kodama,
\newblock Phys. Lett. {\bf B777}, 369 (2018), arXiv:1707.09878.

\bibitem{phsd-01}
Y.~Xu {\em et~al.},
\newblock Phys. Rev. {\bf C96}, 024902 (2017), arXiv:1703.09178.

\end{thebibliography}

\end{document}